\documentclass[astrosymb]{aastex701}

\usepackage{amsmath,amssymb,amsthm}
\usepackage{physics}

\newcommand{\mb}{\boldsymbol}
\newcommand{\bs}{\boldsymbol}
\newcommand{\phat}{\hat{\boldsymbol{\phi}}}
\newcommand{\rhat}{\hat{\mathbf{r}}}
\newcommand{\zhat}{\hat{\mathbf{z}}}
\renewcommand{\curl}[1]{\nabla\times \mb{#1}}
\newcommand{\delt}{\partial_t}
\newcommand{\obar}{\bar{\omega}}
\newcommand{\muo}{\mu_0}

\begin{document}

\title{Global Non-Axisymmetric Hall Instabilities in a Rotating Plasma}

\author[orcid=0009-0001-2520-5366,sname='Alexandre Sainterme']{A.P. Sainterme}
\affiliation{Department of Astrophysical Sciences, Princeton University, Princeton, NJ 08544, USA}
\email[show]{as2252@princeton.edu}

\author[orcid=0000-0003-3109-5367]{Fatima Ebrahimi}
\affiliation{Department of Astrophysical Sciences, Princeton University, Princeton, NJ 08544, USA}
\affiliation{Princeton Plasma Physics Laboratory, Princeton, NJ 08540, USA}
\email{febrahim@pppl.gov}

\begin{abstract}
Non-axisymmetric, flow-driven instabilities in the incompressible Hall-MHD model are studied in a differentially rotating cylindrical plasma.
It is found that in the Hall-MHD regime, both whistler waves and ion-cyclotron waves can extract energy from the flow shear, resulting in two distinct branches of global instability. The non-axisymmetric whistler modes grow significantly faster than non-axisymmetric, ideal MHD modes.
A discussion of the global whistler instability mechanism is presented in the large-ion-skin-depth, `electron-MHD' limit.  When the magnetic field is azimuthal, a subset of the whistler modes having zero axial wave number are uncovered to be destabilized by the `co-rotation amplifier' mechanism. It is observed that the effect of the Hall term on the non-axisymmetric modes can be appreciable when $d_i$ is on the order of a few \% of the width of the cylindrical annulus.  Distinct global modes emerge in the strong Hall-MHD regime at significantly stronger magnetic fields than those required for unstable global MHD modes, as the Hall effect weakens the stabilizing `field-line bending' by decoupling ion motion from the magnetic field. 
These global non-axisymmetric modes may play an important role in weakly ionized accretion disks. 
\end{abstract}

\section{Introduction} Weakly magnetized, differentially rotating plasmas are unstable to flow-driven hydromagnetic instabilities~\citep{Velikov1959,Chandrasekhar1960}. The relevance of this mechanism to accretion disks was explored in the seminal work of \citet{Balbus1991}, wherein the phenomenon was given the name magnetorotational instability (MRI). The study of these linear instabilities is motivated by the question of whether their onset leads to the generation of turbulence, and if the associated turbulent viscosity contributes significantly to the rate of angular momentum transport in accretion disks that is required by $\alpha$-disk models~\citep{Shakura1973}. While the MRI descriptions of \citet{Velikov1959}, \citet{Chandrasekhar1960}  and \citet{Balbus1991} use ideal magnetohydrodynamics (MHD) to model the ionized gases in an accretion disk, it has been noted that plasma behavior beyond ideal MHD is likely important in weakly ionized protoplanetary disks (PPDs)~\citep{Blaes1994}. Other work~\citep{Wardle1999,Balbus2001,Kunz2004,Ebrahimi2011} has specifically considered the influence of the Hall term, $\mb{J}\times\mb{B}/en_e $, on the linear dispersion relation for the axisymmetric MRI using local, WKB-like approximations of the governing differential equations. Linear studies of MRI modes in the non-ideal regime have shown that inclusion of the Hall term modifies the axisymmetric dispersion relation in two ways. First, \citet{Balbus2001} note that Hall effect allows instability in systems that have a radially increasing rotation rate. Increasing rotation profiles are predicted to be stable in ideal MHD. The second observation is that the local dispersion relation becomes asymmetric with respect to the orientation of the magnetic field relative to the axis of rotation/vorticity of the flow. Local linear theory was developed by \citet{Kunz2008} for a non-rotating flow instability in Hall-MHD using a shearing box approximation that was dubbed the Hall-shear instability (HSI). The flow shear driven Hall-MHD instability in non-rotating plasmas has also been referred to as a magnetoshear instability (MSI)~\citep{Bejarano2011}. 3D Hall-MHD calculations show that the MSI can out-compete the growth of Kelvin-Helmholtz instabilities in a shear layer with perpendicular magnetic field if the vorticity is sufficiently large~\citep{Gmez2014}. It has also been shown that there exists a second branch of instability in the local Hall-MHD dispersion relation that results from coupling of the ion-cyclotron wave to the epicyclic motion of a differentially rotating disc~\citep{Simon2015}. In these local linear calculations~\citep{Wardle1999,Balbus2001,Kunz2008,Bejarano2011,Simon2015}, it was shown that growth rate depends on the relative orientation of the vorticity and the magnetic field, and instability is possible for either positive or negative shear (increasing or decreasing rotation rate).

\par Recent work solving the radial eigenvalue problem has elucidated the importance of global non-axisymmetric modes in the MHD model due to the cylindrical curvature of the geometry and flow curvature~\citep{Ebrahimi2022,Ebrahimi2025,Haywood2025acc}, as well as in laboratory settings~\citep{wang2025observation}. Alfv\'enic resonances~\citep{matsumoto1995magnetic,Ogilvie1996,Ebrahimi2022} are shown to be instrumental in the global nature of the MHD non-axisymmetric flow-driven and curvature modes (the global non-axisymmetric MRI branch). The question arises whether dispersive Hall-MHD modes contribute to the onset of global flow-driven instabilities in differentially rotating systems. Here, we uncover that in fact whistler and ion-cyclotron waves can have a dominant effect on the onset of non-axisymmetric modes in the Hall-MHD model. 
\par This work presents a linear study of the spectrum of non-axisymmetric, flow driven instabilities in the Hall-MHD model in a rotating plasma Couette flow of the type one might expect near the mid-plane of a protoplanetary accretion disk. In the Hall-MHD model, we show that there are two distinct branches of instability in the presence of the Hall effect. One branch is faster growing than the MHD MRI modes at modest values of the ion skin depth, $d_i$. We identify this branch as a  whistler wave that is able to extract energy from the steady-state flow shear. The second branch is akin to the incompressible MHD modes with growth rates and mode structures that are modified by the Hall term. We also explore the instabilities of the rotating system in the large $d_i$, electron-MHD (EMHD) limit. In this regime, ions provide an effectively stationary, neutralizing background through which the electron fluid moves. This model retains the coupling between the mean rotation profile and the whistler waves. In this way, the EMHD model isolates the whistler branch of linear instability that is not simply explained as a destabilizing effect on the non-axisymmetric MHD MRI modes. Both local and global analysis and numerical calculations of the electron MHD system are presented. 

Using a combination of local and global eigenvalue analysis as well as numerical calculations, we assess the degree to which the Hall-MHD waves extract energy from the differential rotation - leading to new non-axisymmetric modes that exist at larger magnetic field strengths than MHD modes~\citep{Ebrahimi2022}. We find that for purely parallel propagating modes in the EMHD model, the local WKB theory predicts stability despite the presence of global unstable whistler modes.  

\par The organization of this paper is as follows. Section \ref{sec: hmhd} introduces the Hall-MHD model and investigates incompressible, non-axisymmetric instabilities in differentially-rotating cylindrical disk geometry. We present numerical results using two different methods. The strength of the hall term is varied by adjusting the ion skin depth, $d_i=c/\omega_{pi}$, where $\omega_{pi}\equiv \sqrt{Z^2e^2n_i/\epsilon_0m_i}$ is the ion plasma frequency. Section \ref{sec: emhd} presents analysis and numerical calculations of the EMHD limit. It is shown that non-axisymmetric whistler waves are driven unstable by the sheared rotation when the wave frequency is similar in magnitude to the rotation rate. The local, point-wise dispersion relation theory is compared to the results of global analysis. Necessary conditions for the existence of unstable modes are derived based on global properties of the EMHD system. Numerical results for the global eigenvalue problem are presented at several values of disk aspect ratio (height to width), in an attempt to study the role of the boundary locations on global whistler instabilities. It is found that lower aspect ratio (thinner disk) generally leads to instability over a broader range of magnetic field strengths. 

\section{The Hall-MHD model}
\label{sec: hmhd}
We consider an incompressible Hall-MHD model of plasma dynamics. The full system of nonlinear PDEs couples the bulk fluid motion of a quasi-neutral two-component plasma to `pre-Maxwell' classical electromagnetism. The equations describing the fluid motions include an equation of continuity for the plasma mass density, $\rho$, moving with velocity field $\bs{v}$.
\begin{equation}
    \delt \rho + \nabla\cdot\left(\rho\bs{v}\right) = 0.
    \label{eq: contintuity}
\end{equation}
    
Assuming electron inertia is negligible, the equation governing the total momentum evolution is
\begin{equation}
    \rho\delt \bs{v} + \rho\bs{v}\cdot\nabla\bs{v} = \frac{\bs{B}\cdot \nabla\bs{B}}{\mu_0} - \nabla P - \rho\nabla\Phi_g.
    \label{eq: momentum}
\end{equation}
Here, $P$ is the sum of the isotropic gas pressure, $p$, and the magnetic pressure, $B^2/2\mu_0$. $\Phi_g$ is the Newtonian gravitational potential. We assume the source of the gravitational potential is a central body of mass $M$, and neglect self-gravitation of the disk material, so that $\Phi_g = -GM/r$. The time-dependence of the magnetic field $\mb{B}$ is determined by Faraday's law, 
\begin{equation}
    \delt \mb{B} = -\nabla\times\mb{E},
    \label{eq: induction}
\end{equation}
and the electric field is given by the Hall-MHD Ohm's law,
\begin{equation}
    \bs{E} = -\bs{v}\times\bs{B} + \frac{\mb{J}\times\mb{B}}{en_e}.
    \label{eq: ohms}
\end{equation}
Here, $n_e$ is the electron density, which can be approximately related to the mass density, $\rho$ by $n_e\sim Z\rho/m_i$ for a two-component plasma with ions with charge $Ze$, and $m_e\ll m_i$. In the MHD approximation, we neglect displacement current, so $\mu_0\bs{J}=\nabla\times\bs{B}$. The system of equations is completed by incompressibility, $\nabla\cdot\mb{v}=0$, and the magnetic divergence constraint $\nabla\cdot\bs{B}=0$. 
\par
\begin{figure}
    \centering
    \includegraphics[width=0.4\linewidth]{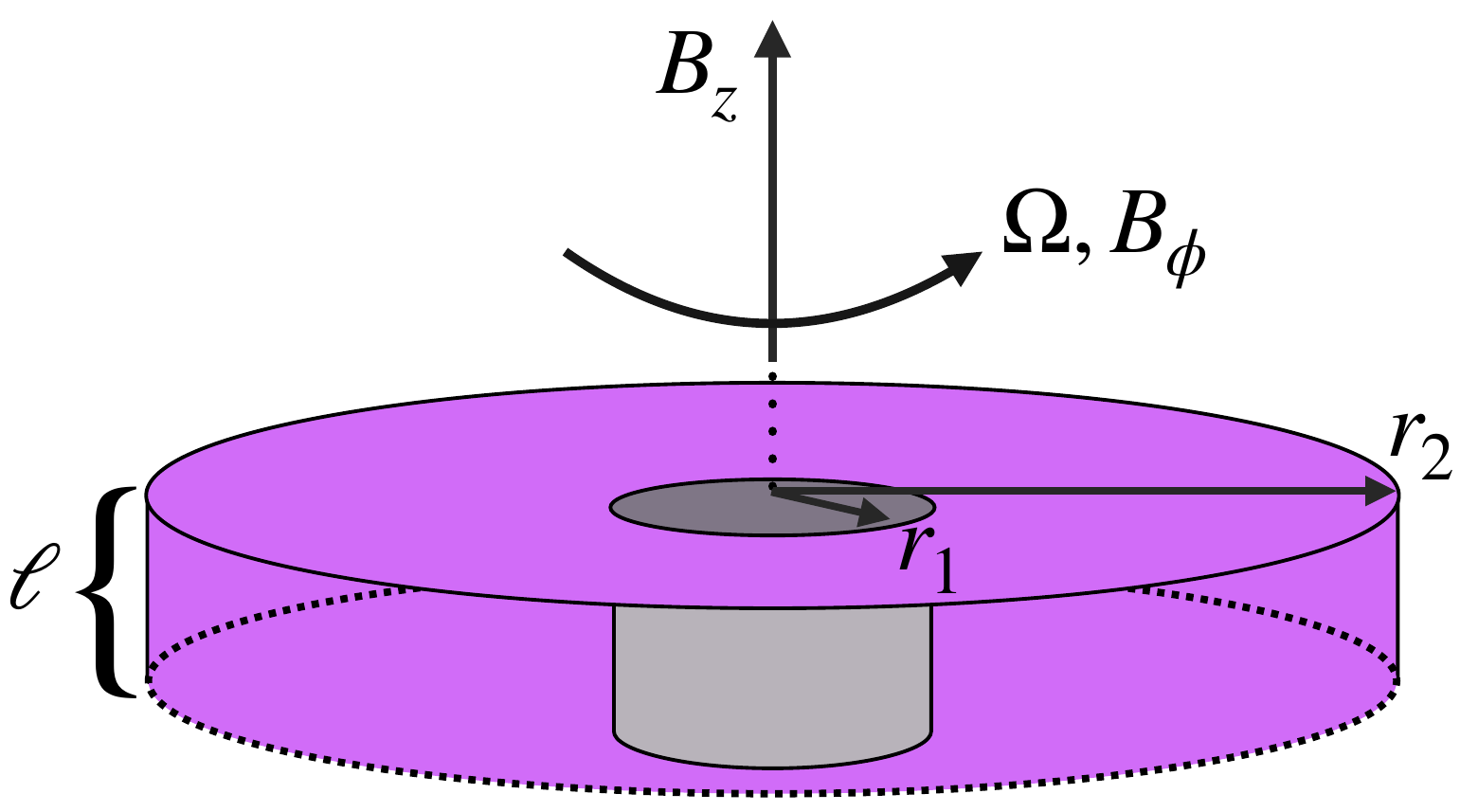}
    \caption{Schematic representation of the cylindrical rotating plasma equilibrium.}
    \label{fig: eq-schem}
\end{figure}

 To address the question of linear stability, the equations are linearized around a steady-state, equilibrium solution in cylindrical $(r,\phi,z)$ coordinates (see figure \ref{fig: eq-schem}). The steady-state is defined by an axisymmetric, rotating, uniform-density, current-free azimuthal flow ($\bs{v}=r\Omega\phat$) that is supported by either pressure gradients or a gravitational potential. The cylindrical domain is intended to represent an annular slice of the accreting material very near the mid-plane of the disk. We assume that the vertical domain is small enough that we may neglect stratification of the plasma and the vertical components of $\nabla \Phi_g$. We also assume azimuthal symmetry of the steady-state. Thus, variation of the steady-state fields in the direction away from the disk mid-plane and in the azimuthal direction is neglected; $\partial_\phi Q = \partial_z Q=0$ for any $Q\in \{\Omega,\bs{B},\Phi_g,P,\rho\}$. The remaining non-trivial relations for the steady-state are
\begin{equation}
    \partial_rB_z=\partial_r(rB_\phi) = 0,
\end{equation}
and
\begin{equation}
    \rho r\Omega^2 = \partial_r P + \rho \frac{GM}{r^2}.
\end{equation}
    If $\partial_rP=0$, $\Omega = \sqrt{GM/r^3} $ is the Keplerian rotation rate -- the orbital frequency of circular orbits in the gravitational potential created by the central body. Arbitrary $\Omega(r)$ profiles can be accommodated in the steady-state by including a nonzero radial pressure gradient. In the incompressible case considered in this work, the linearized equations are unaffected by the steady-state radial pressure gradient.

\par  To study linear stability of the steady flow, we linearize the equations by perturbing each quantity by a small amount. The magnetic field, for example, becomes $\bs{B}+\bs{b}$. Here, $\bs{B}$ is a solution to the steady state described above, and we assume $|\bs{b}|/|\bs{B}| \ll 1$. Applying this procedure to each variable in equations \eqref{eq: contintuity}-\eqref{eq: ohms}, and neglecting any terms that depend on higher powers of perturbed quantities, one arrives at the following incompressible linearized Hall-MHD equations for a fully-ionized plasma linearized about a steady, differentially rotating azimuthal flow are given, in cylindrical coordinates, by
\begin{equation}
    \rho\partial_t\mb{v} + \Omega\partial_\phi \rho\mb{v} +\left(\frac{\kappa^2}{2\Omega}\phat\rhat -  2\Omega\rhat\phat\right)\cdot\rho\mb{v} = -\nabla \tilde{P} + \frac{1}{\mu_0}\left(\mb{B}\cdot\nabla\mb{b} + \mb{b}\cdot\nabla\mb{B}\right),\label{eq: basic-mom}
\end{equation}
\begin{equation}    
      \partial_t\mb{b}+\Omega\partial_\phi\mb{b}-r\Omega'\phat b_r = \mb{B}\cdot\nabla\mb{v}+\frac{2B_\phi}{r}v_r\phat + \frac{d_i}{\sqrt{\mu_0\rho}}\left(\curl{b}\cdot\nabla\mb{B}- 
      \mb{B}\cdot\nabla\curl{b}\right),\label{eq: basic-amp}
\end{equation}
\begin{equation}
      \nabla\cdot\mb{v} =\nabla\cdot\mb{b} =  0 \label{eq: basic-div}.
\end{equation}

 $\kappa^2 \equiv 4\Omega^2 + 2\Omega r\Omega'$ is the epicyclic frequency, $\mb{B}$ is the background magnetic field, and $d_i=c/\omega_{pi}$ is the ion skin depth. The lowercase variables $\mb{v},\mb{b}$ represent the perturbed flow and magnetic field respectively, and $\tilde{P}=p+\mb{B}\cdot\mb{b}/\mu_0$ is the perturbed total pressure. We note that the identification $1/en_e\rightarrow d_i/\sqrt{\mu_0\rho}$ is only valid to lowest order in $m_e/m_i$, and assumes $\rho \approx n_em_i$. If one were to consider a weakly-ionized plasma where $\rho$ and $\bs{v}$ correspond to the total bulk density and flow including neutral gas, then $d_i$ ought to be replaced by $\ell_H\equiv \sqrt{(\rho/\rho_i)} d_i $~\citep{Pandey2008}.
\par  The axial component of the magnetic field, $B_z$, is presumed uniform throughout the domain, and the azimuthal component satisfies the current-free condition; $\partial_r (rB_\phi) = 0$. No assumption is made regarding the specific form of $\Omega(r)$ while deriving the following equations. However, since it is anticipated than Hall effects are most dominant in protoplanetary accretion disks~\citep{Wardle2012}, we will consider the case of Keplerian rotation where $\Omega \propto r^{-3/2}$ when computing numerical values for the growth rates.
\par Following the work of \citet{Chandrasekhar1960} and \citet{Frieman1960} in the ideal MHD case, it is convenient to introduce the Lagrangian displacement $\bs{\xi}$ such that $\mb{v} = \partial_t\bs{\xi} + \bs{V}\cdot\nabla\bs{\xi} - \bs{\xi}\cdot\nabla\bs{V}$. The ideal MHD system can be rewritten as a single second-order ODE in the radial component of the displacement $\xi_r$~\citep{Khalzov2006,Ebrahimi2022,Ebrahimi2025}.
If the perturbations are axisymmetric, $\partial_\phi \bs{\xi}=\partial_\phi{\bs{b}}=0$, the resulting ODE is an eigenvalue problem for a Hermitian differential system whose eigenvalues are $\omega^2$~\citep{Chandrasekhar1961}. In that case, the problem of determining stability of the system is reduced to finding parameters for which a solution of the system of ODEs satisfying given boundary conditions has a negative eigenvalue. In the present case we do not assume axisymmetry, so the problem cannot be reduced to finding the eigenvalues of a Hermitian differential operator. Furthermore, the inclusion of the Hall term complicates the system so that reduction to a single second order equation in $\xi_r$ is not possible. However, we elect to write the equations in terms of $\bs{\xi}$ for the sake of comparison. 
\par Considering perturbations of the form $\mb{\xi},\mb{b}\propto \exp(im\phi + ikz - i\omega t)$, we arrive at a set of coupled ordinary differential equations in the radial coordinate that define an eigenvalue problem for the complex frequency $\omega$. After substituting the complex exponential ansatz, three of the resulting equations are reduced to algebraic relations between components. Eliminating the purely algebraic constraints, one can derive a system of four coupled first-order ODEs in the radial coordinate. Let $\psi\equiv irb_r,\varphi\equiv rb_\phi$, and $\chi\equiv r\xi_r$. Then, the linearized system of incompressible Hall-MHD equations, eqs. \eqref{eq: basic-mom}-\eqref{eq: basic-div}, can be written in the following form after eliminating $\xi_\phi,\xi_z$ and $b_z$.
\begin{equation}
		\begin{split}
        	(\kappa^2-\obar^2)F\chi + rF\tilde{P}' - \frac{2\Omega}{\obar}mF\tilde{P} = \left(1 - \frac{2\Omega}{\obar}\frac{B_\phi}{rF}\right)\omega_A^2\psi -\left(\frac{2\Omega}{\obar} + \frac{B_\phi}{rF}\right)\omega_A^2\varphi  ,
	 \label{eq: hmhd-1}
\end{split}
\end{equation}
\begin{equation}
		\obar^2rF\chi' + 2\Omega\obar m F\chi + \omega_A^2\left(r\psi'  + \frac{mB_\phi}{rF}\psi\right) = (m^2+r^2k^2)F\tilde{P} ,
\label{eq: hmhd-2}
\end{equation}
\begin{equation}
	r^2k\obar( \psi + F\chi ) = -  V_H(mr\psi' - (m^2+r^2k^2)\varphi ), \label{eq: hmhd-3}
\end{equation}
\begin{equation}
	r^2k^2F\tilde{P} = (\omega_A^2-\obar^2)(r \psi'-m\varphi) +  kV_H\obar(r\varphi'-m\psi). \label{eq: hmhd-4}
\end{equation}
Here, $\obar\equiv \omega-m\Omega(r)$, $F\equiv \mb{k}\cdot\mb{B}$, $\omega_A \equiv F/\sqrt{\mu_0\rho}$, and $V_H\equiv \omega_Ad_i$. $\omega_A$ is the local frequency of a shear Alfv\'en wave, and $V_H$ is the phase velocity of a parallel propagating whistler wave. Compared to the linear, incompressible, ideal MHD equations, equations \eqref{eq: hmhd-1}-\eqref{eq: hmhd-4} include two additional first-order differential equations in the radial coordinate that complicate the relationship between the magnetic field and fluid motion. In particular, we note that the radial components of the perturbed displacement and magnetic field are no longer simply proportional to each other. The additional terms in equations \eqref{eq: hmhd-3} and \eqref{eq: hmhd-4} reflect the fact that in non-dissipative Hall-MHD, the magnetic field is `frozen in' to the electron fluid motion instead of the bulk ion motion~\citep{Hameiri2005}. This is in contrast to the ideal MHD case, where equation \eqref{eq: hmhd-3} simplifies to $\psi = -F\chi$.
\par
\begin{figure}[htb]
    \centering
    \includegraphics[width=0.5\linewidth]{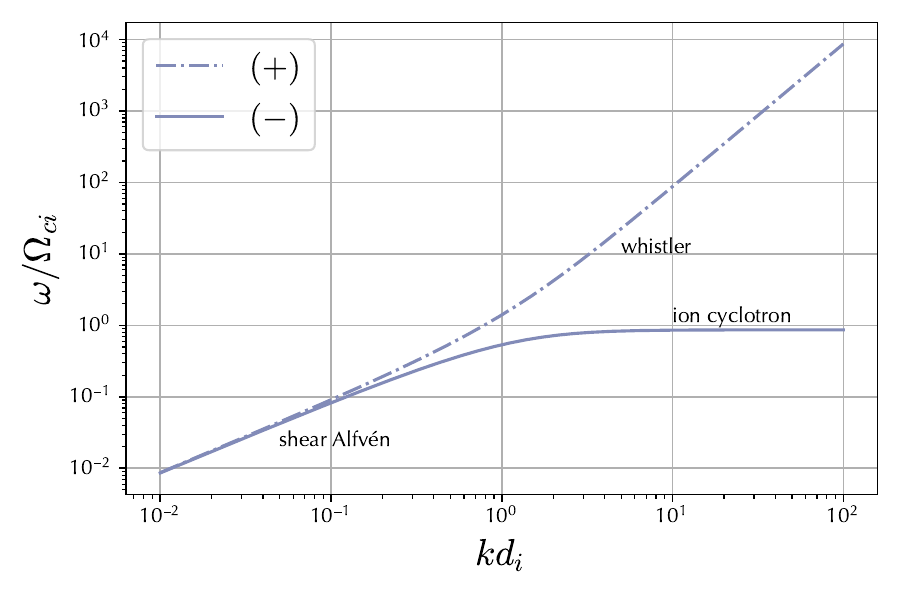}
    \caption{Plot of the positive solutions of $(\omega^2-\omega_A^2) =  \pm kd_i \omega \omega_A$ as a function of $kd_i$. $\Omega_{ci}\equiv |V_A|/d_i$ is the ion cyclotron frequency, $V_A=B/\sqrt{\muo\rho}$ is the Alfv\'en velocity and $\omega_A\equiv k_\parallel B/\sqrt{\mu_0\rho}$ is the associated frequency of shear-Alfv\'en waves. Here, we choose nearly parallel propagation: $k_\parallel=k\cos(\pi/6)$}
    \label{fig: hmhd_incomp_disp}
\end{figure}

 To orient our discussion of the Hall effect on the spectrum of instabilities present in the system, it is useful to first consider the dispersion relation for a magnetized, homogeneous plasma in the incompressible Hall-MHD limit. It possesses two sets of roots: $(\omega^2-\omega_A^2)^2 =  k^2d_i^2 \omega^2 \omega_A^2$. The two solutions of this dispersion relation are plotted as a function of $kd_i$ in figure \ref{fig: hmhd_incomp_disp}. The pair of solutions with the larger phase velocity describe whistler waves when $kd_i$ is large, and are analogous to the MHD fast magnetosonic wave in the compressible case at low-$kd_i$. The slower solution for $\omega^2$ is the Hall-modified shear Alfv\'en wave in the low-$kd_i$ regime, which transitions into ion cyclotron waves propagating parallel to $\mb{B}$ in the large $kd_i$ limit\footnote{In the incompressible case, the slow wave and shear Alfv\'en wave are degenerate. The shear Alfv\'en mode is incompressible at low $kd_i$, but the slow wave branch is incompressible at large $kd_i$.}~\citep{Hameiri2005}. In the incompressible limit, the fast wave and the shear Alfv\'en wave are degnerate at $kd_i\ll 1$ (see figure \ref{fig: hmhd_incomp_disp}). However, as we will show in Section \ref{sec: hmhd-numeric}, small but non-zero $d_i$ lifts the degeneracy between the two wave branches, leading to qualitatively different behavior in the differentially rotating flow. 
One interesting question is whether this taxonomy of waves in a uniform, magnetized plasma can be simply related to the spectrum of instabilities in the differentially rotating cylinder. To this end, in Section \ref{sec: emhd} we reduce the full linear eigenvalue problem into subsystems that represent the different mode branches of interest. For the following presentation of numerical results from the Hall-MHD model, it suffices to appreciate that the strength of Hall effect on linear modes is chiefly controlled by the parameter $kd_i$.

\subsection{Numerical results for Hall-MHD stability }
\label{sec: hmhd-numeric}
 In this section we provide two types of numerical solution of the linearized Hall-MHD equations \eqref{eq: basic-mom}-\eqref{eq: basic-div}. The first approach uses a spectral element method to compute the eigenvalues of the incompressible Hall-MHD system in a periodic cylinder. For a given $m,k$ pair, we write $\bs{v},\bs{b}\propto \exp(im\phi+ikz-i\omega t)$, and the radial dependence of the solutions is represented using a basis of piecewise differentiable polynomials (typically cubic or quartic). Projecting the equations onto this basis results in a linear eigenvalue problem in terms of the vector of coefficients of the expansion that is solved using the LAPACK ZGGEV routine~\citep{laug}. The problem is formulated so that the resulting eigenvalue is the complex frequency describing the time-dependence of the solution: $\omega = \omega_r + i\gamma$. Incompressibility is enforced using an auxiliary variable that is algebraically eliminated in a manner consistent with a penalty method. The magnetic divergence constraint is enforced with a diffusive divergence error correction term. The code we employ is a version of the CYL\_SPEC code \citep{Sovinec2016} modified to included the Hall term and sheared rotation in the equilibrium.
\par We compare a subset of the eigenvalue results with linear initial value calculations using the extended MHD code NIMROD~\citep{Sovinec2004}. NIMROD (non-ideal magnetohydrodynamics with rotation, open discussion), uses high-order spectral element method to solve both linear and nonlinear extended MHD. The domain is discretized using a 2-D grid of high-order polynomial finite elements in the $r-z$ plane, and a pseudospectral collocation method with a Fourier basis for the periodic azimuthal coordinate, $\phi$. The equations are integrated in time using a semi-implicit leapfrog time-stepping scheme. The accuracy and numerical stability of the two-fluid model in NIMROD has been successfully benchmarked with simulations of linear and nonlinear non-ideal MHD instabilities~\citep{Sovinec2010}. 
\par In order to compare NIMROD calculations to CYL\_SPEC, we compute an effective growth rate from the time-series data output of the code. For a given azimuthal mode number, the velocity and magnetic field are initialized with a small amplitude perturbation. The linear equations are integrated in time until the growth of the most unstable mode can be unambiguously identified. The linear growth rate, $\gamma$, is computed by calculating the change in either the perturbed magnetic and kinetic energies between successive time steps. Concretely, $2\gamma = \log(\Delta \int b^2/2\mu_0~d^3x / \Delta t)$ or $2\gamma = \log(\Delta \int \rho v^2~d^3x / \Delta t)$, where $b,v$ are the magnitudes of the perturbed magnetic field and perturbed flow, respectively. 
\par We note that the eigenvalue solver does not include electrical resistivity or viscosity. In the NIMROD calculations, however, a small amount of resistivity and viscosity is used for improved numerical stability. The effect of these dissipative terms on the linear growth of the axisymmetric MRI, and non-axisymmetric `Magneto-Curvature modes' modes in MHD is understood \citep{Balbus1998, Ebrahimi2025,Haywood2025acc}. It is also well-established that there can be a competition between the resistive dissipation and the Hall term \citep{Wardle2012}. The value of magnetic Reynolds number in the NIMROD calculations is $\mathrm{Rm}=2529$, and the magnetic Prandtl number is $\mathrm{Pm}=1$. With these parameters, the effect on the most unstable linear modes is negligible. In the parlance of \citet{Wardle2012}, we are firmly in the `Hall-MRI' regime. In this section we focus on the most global (i.e. lowest $m,k$) modes, since these modes would be the least affected by dissipative mechanisms. To assess the effect of varying strength of the Hall term, we compare the results for different values of $d_i$ to both the MHD result, and the EMHD limit with $d_i=1r_1$ for both azimuthal and axial magnetic field configurations in Sections \ref{sec: hmhd-axial} and \ref{sec: hmhd-azimuthal}, respectively. For the calculations in this section, we normalize lengths by the inner wall radius of the domain, $r_1$, and set $r_2=5r_1$, $\ell = 8r_1$. Times are normalized by $\Omega_0\equiv\Omega(r_1)$, and velocities by $r_1\Omega_0$. Since density is uniform, we give magnetic field strengths in terms of the corresponding normalized Alfv\'en velocities, $V_A/r_1\Omega_0$. This is also referred to as the Lehnert number~\citep{Lehnert1954}

\subsection{Axial Magnetic Field}
\label{sec: hmhd-axial}
\par
\begin{figure}[htbp]
    \centering
    \includegraphics[width=0.5\linewidth]{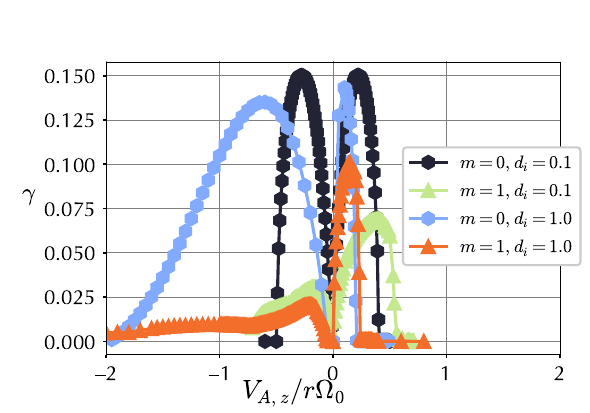}
    \caption{Growth rates of the $m=0$ and $m=1$ modes with $k=\pi/4$ with $d_i=0.1$ and $d_i=1.0$. The Hall stronger hall term creates a notable asymmetry between $B_z>0$ and $B_z<0$.}
    \label{fig: hmhd_m0}
\end{figure}

\begin{figure}[htbp]
    \centering
    \includegraphics[width=0.9\linewidth]{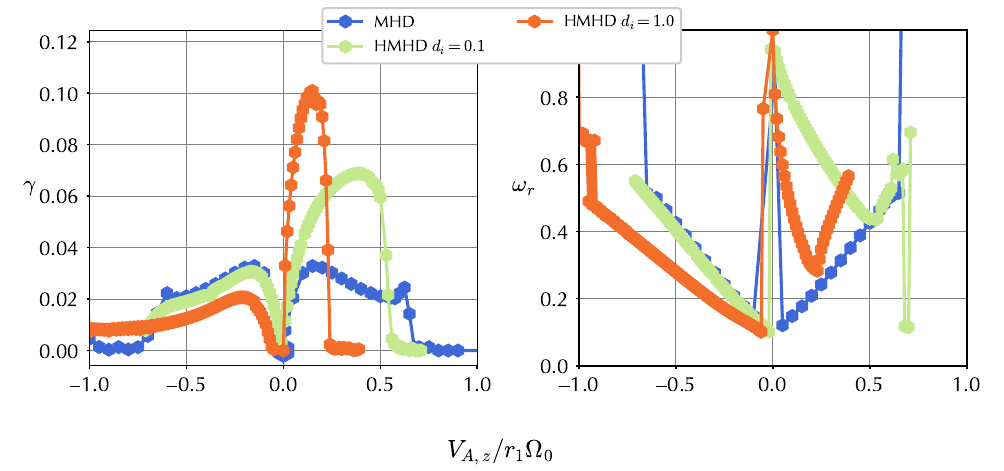}
    \caption{Growth rates and frequencies of the most unstable mode with $m=1,k=\pi/4$ as a function of the axial magnetic field strength measured by the inverse Alfv\'enic Mach number. We compare the MHD case in blue to the Hall-MHD result at $d_i=0.1$, and $d_i=1$ in green and orange, respectively. As the strength of the  Hall effect ($d_i$) increases, the growth rate develops an asymmetry with respect to the direction magnetic field.}
    \label{fig: hall_mhd_vertical_m1_k1}
\end{figure}

 We first consider the case in which the magnetic field is oriented axially, so $\bs{B}=B_z\zhat$. The linear growth of the axisymmetric, $m=0$ modes in Hall-MHD has been studied~\citep{Wardle1999,Ebrahimi2011,Wardle2012}. Here, we present the linear Hall-MHD results obtained for m=1 modes using the eigenvalue code (CYL\_SPEC). We find that it is generally the case that non-axisymmetric modes with $m=1$ are the fastest growing at large magnetic field strengths, but are subdominant to axisymmetric modes at lower magnetic field strengths. The exact relationship depends on both the orientation of the field relative to the direction of rotation, and the value of $d_i$. For completeness, we compare the growth rates of the axisymmetric $m=0$ to the non-axisymmetric $m=1$ modes at two values of $d_i$ as a function of $V_{A,z}/r_1\Omega_0$ in figure \ref{fig: hmhd_m0}. 
For both values of $d_i$ the $m=0$ mode has a larger growth rate when $|B_z|$ is small, and a higher maximum growth rate. There is an intermediate value of $|B_z|$, however, where the $m=0$ mode is stable, and the $m=1$ is the dominant mode. We do not plot the mode frequencies here, since the $m=0$ has $\omega_r=0$, and the frequency of the $m=1$ modes are shown in figure \ref{fig: hall_mhd_vertical_m1_k1}. 
\par Figure \ref{fig: hall_mhd_vertical_m1_k1} plots the growth rates and frequencies of the fastest growing non-axisymmetric, $m=1$ modes as a function of the magnetic field strength for the MHD model and two values of $d_i$ in the Hall-MHD model. The effect of the Hall term is significant even at modest values of $d_i$. At $d_i=0.1r_1$, the ion skin depth is only 2.5\% of the radial extent of the disk, and $kd_i\approx 0.08$, yet the maximum growth rate of the fastest growing mode is more than twice as large as in the MHD case. The axial field strength at which the maximum growth rate is attained is also larger.
\par The Hall term destabilizes a different, high-frequency ($\omega \sim \Omega$) branch of non-axisymmetric instability that grows significantly faster than either the $m=1$ MRI or the low frequency global curvature mode in the MHD limit~\citep{Ebrahimi2022}.
As $d_i$ is increased, the dispersion relation develops an increasingly asymmetric dependence on direction of $\mb{B}$ relative to the axis of rotation as the system transitions from the MHD to the Hall-dominant regime. In the right-handed cylindrical coordinates used, we assume $\bs{\Omega}\propto \zhat$, so that positive $B_z$ corresponds to a magnetic field parallel to the axis of rotation. As $d_i$ increases, the $m=1,k=1$ modes with $B_z<0$ have lower growth rates than the modes with $B_z>0$, but remain unstable at larger values of $|B_z|$.

\par
\begin{figure}[htb]
    \centering
    \includegraphics[width=0.32\linewidth]{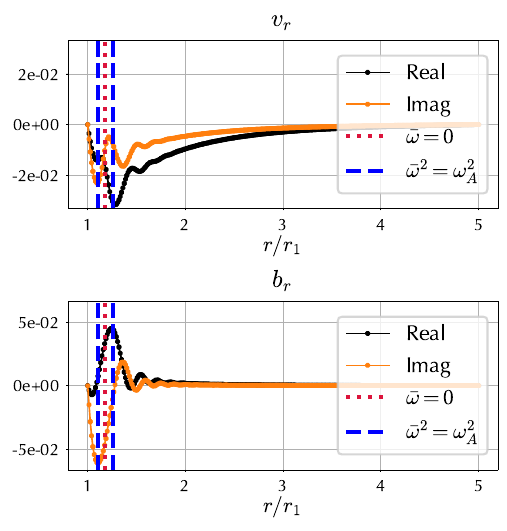}\;~\includegraphics[width=0.32\linewidth]{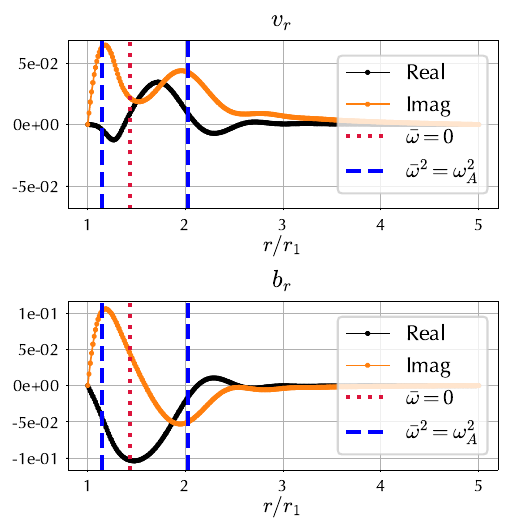}\;~\includegraphics[width=0.32\linewidth]{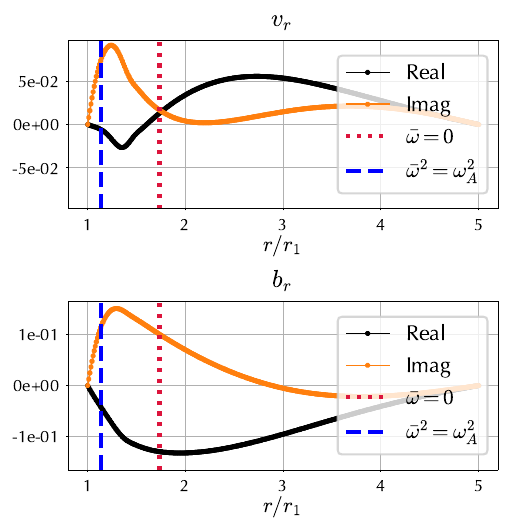}
    \caption{Radial structure of eigenfunctions for Hall-MHD system with $d_i=0.1 r_1$ and axial magnetic fields $V_{A,z}=0.1r_1\Omega_0$ (left), $V_{A,z}=0.3r_1\Omega_0$ (center)
    and $V_{A,z}=0.5r_1\Omega_0$ (right)}
    \label{fig: hmhd_eigprofs}
\end{figure}

 Figure \ref{fig: hmhd_eigprofs} shows the radial profiles of the solutions to the Hall-MHD eigenvalue problem at $d_i=0.1$ for $V_{A,z}=0.1r_1\Omega_1$,  $V_{A,z}=0.3r_1\Omega_0$ and $V_{A,z}=0.5r_1\Omega_0$.

As the strength of the axial field decreases, the solutions become increasingly oscillatory. This shortening of the radial wavelength is expected from whistler modes. We investigate this point further in section \ref{sec: emhd}. 

\par
\begin{figure}[htb]
    \centering
    \includegraphics[width=0.32\linewidth]{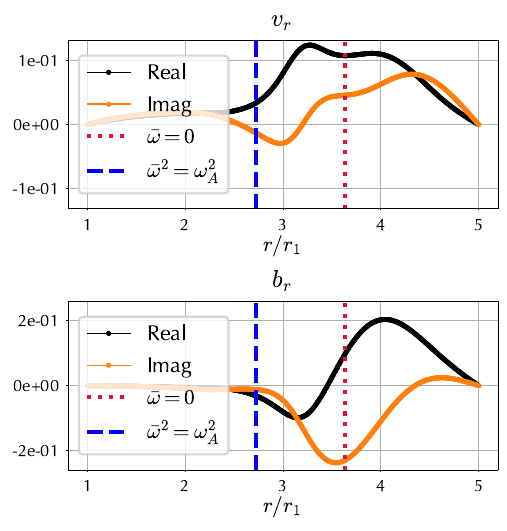}\;~\includegraphics[width=0.32\linewidth]{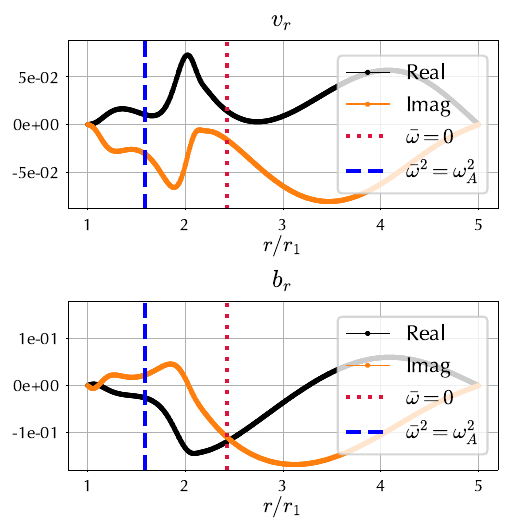}\;~\includegraphics[width=0.32\linewidth]{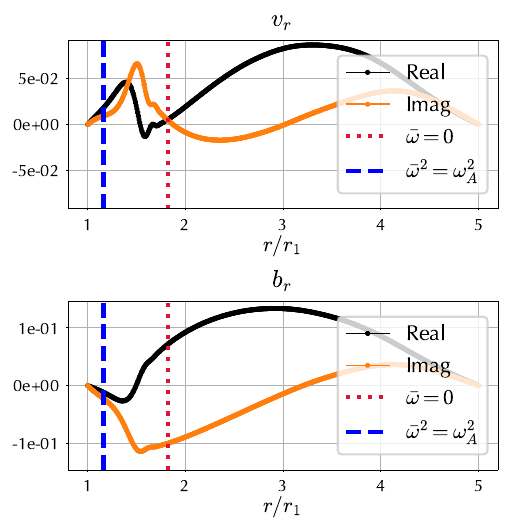}
    \caption{Radial structure of eigenfunctions for Hall-MHD system with $d_i=0.1 r_1$ and axial magnetic fields $V_{A,z}=-0.1r_1\Omega_0$ (left), $V_{A,z}=-0.3r_1\Omega_0$ (center)
    and $V_{A,z}=-0.5r_1\Omega_0$ (right)}
    \label{fig: hmhd_bz<0_eigprofs}
\end{figure}

 When the axial field is anti-parallel with the axis of rotation, $B_z<0$, the whistler modes are not present for $m=1,k=\pi/4$. Instead, the most unstable modes are more similar in radial structure to the low-frequency global `curvature modes' of \citet{Ebrahimi2022} that have been modified by the Hall term. As $d_i$ increases, the growth rates of the curvature modes for $B_z<0$ decrease. Figure \ref{fig: hmhd_bz<0_eigprofs} shows the eigenmodes from the same three $|V_{A,z}|$ as in figure \ref{fig: hmhd_eigprofs}, but with $V_{A,z}<0$. 
\subsection{Azimuthal Magnetic Field}
\label{sec: hmhd-azimuthal}
\par
\begin{figure}[htb]
    \centering
   \includegraphics[width=0.9\linewidth]{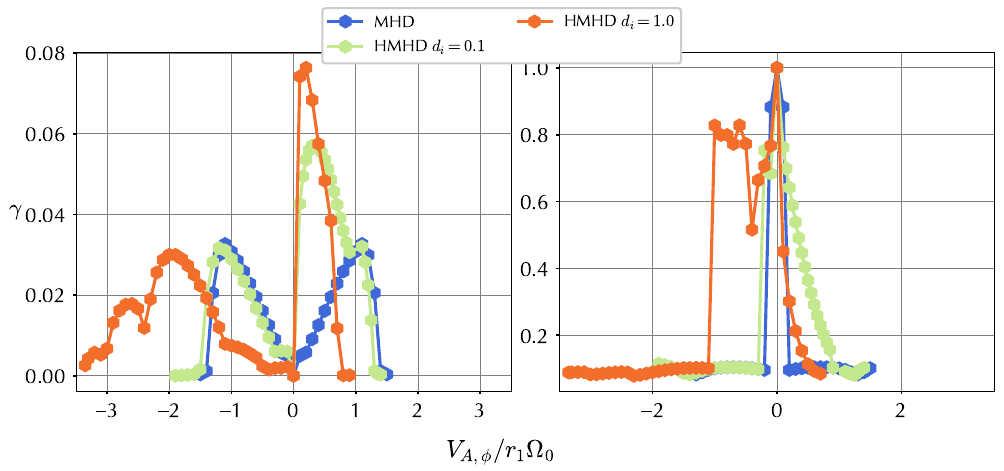}
    \caption{Growth rates and frequencies of the most unstable eigenmode with $m=1,k=\pi/4$ as a function of the azimuthal magnetic field strength at the inner boundary (in units of the inverse Alfv\'enic Mach number). The MHD result (in blue) and Hall-MHD (HMHD) with of $d_i=0.1$ and $d_i=1$ are shown. }
    \label{fig: hall_mhd_azimuthal_m1_k1}
\end{figure}

 When the field is purely azimuthal, $B_\phi\neq0, B_z=0$, the asymmetric effect of the Hall term is more pronounced. Figure \ref{fig: hall_mhd_azimuthal_m1_k1} plots the growth rates and frequencies of the fastest growing $m=1,k=\pi/4$ mode as a function of the azimuthal field strength. When $B_\phi>0$ we see a significant increase in the growth rates of the most unstable non-axisymmetric modes as $d_i$ is increased. Also, with increasing $d_i$, the range of $B_\phi$ over which the modes are unstable is reduced. Figure \ref{fig: hall_mhd_azimuthal_m1_k1} only shows $k>0$ modes, but we note that if $k=-\pi/4$, the resulting dispersion relation is mirrored about the vertical axis.
Based on the behavior of the mode frequencies, we surmise that there are at least two distinct modes present in the calculation. As $|B_\phi|\rightarrow 0$, a higher frequency mode with very slow growth dominates.
\begin{figure}[htbp]
    \centering
    \includegraphics[width=0.32\linewidth]{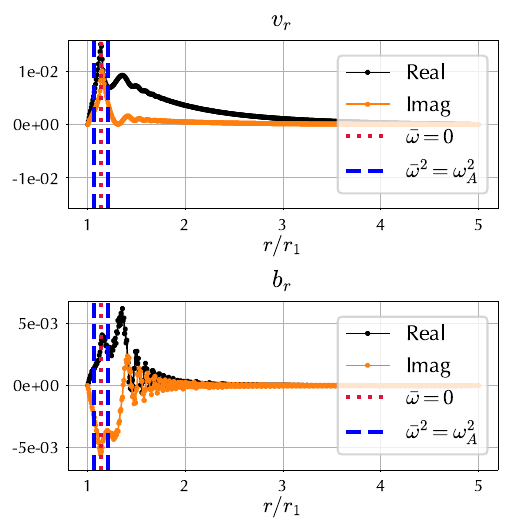}\;~\includegraphics[width=0.32\linewidth]{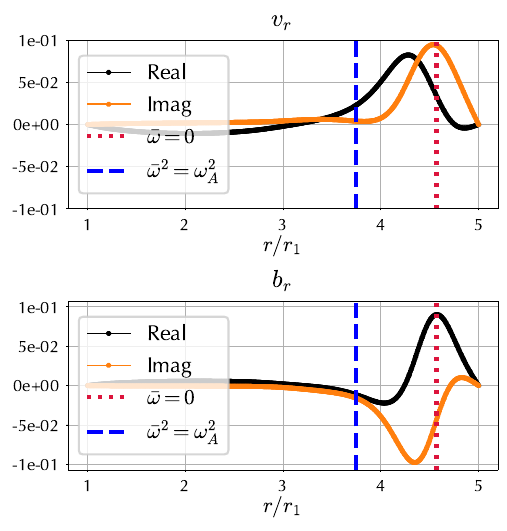}\;~\includegraphics[width=0.32\linewidth]{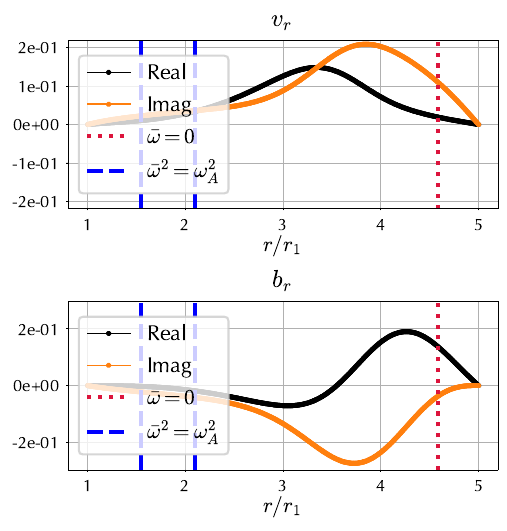}
    \caption{Radial profiles of the $m=1,k=\pi/4$ HMHD eigenfunctions with $d_i=0.1$ with azimuthal fields $V_{A,\phi}=-0.1\Omega_0r_1$ (left), and $V_{A,\phi}=-0.5\Omega_0r_1$ (center), and $V_{A,\phi}=-1\Omega_0r_1$ (right).}
    \label{fig: hmhd_azimuthal_bp<0_profs}
\end{figure}
As $|B_\phi|$ increases, a the lower-frequency mode is destabilized. Unlike the whistler and MRI-type modes, the frequency of this mode is effectively constant as $|B_\phi|$ varies.
\par
\begin{figure}
    \centering
    \includegraphics[width=0.32\linewidth]{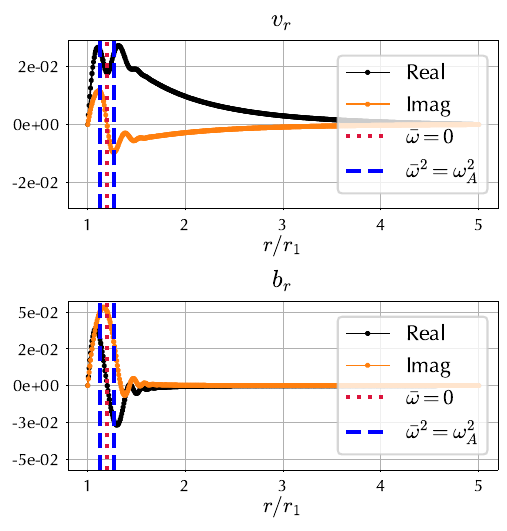}\;~\includegraphics[width=0.32\linewidth]{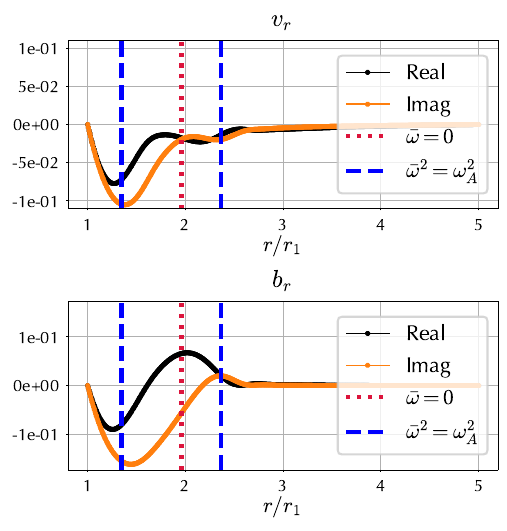}\;~\includegraphics[width=0.32\linewidth]{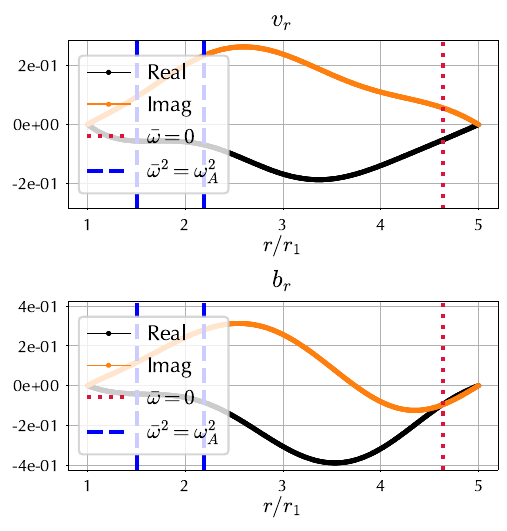}
    \caption{Radial profiles of the $m=1,k=\pi/4$ HMHD eigenfunctions with $d_i=0.1$ with azimuthal fields $V_{A,\phi}=0.1\Omega_0r_1$ (left), and $V_{A,\phi}=0.5\Omega_0r_1$ (center), and $V_{A,\phi}=1\Omega_0r_1$.}
    \label{fig: hmhd_azimuthal_bp>0_profs}
\end{figure}

 The mode structure associated with the lower-frequency Hall-MHD modes that exist at large magnetic field strength are truly global, with non-trivial magnitude extending throughout the domain. Figure \ref{fig: hmhd_azimuthal_bp<0_profs}  shows the mode structures for the $B_\phi<0$ modes, and figure \ref{fig: hmhd_azimuthal_bp>0_profs} shows modes with $B_\phi > 0$.
 The behavior of the low-frequency, large-field modes in the azimuthal HMHD cases suggests that they may be described as the slower Hall-MHD Alfv\'en wave being destabilized by the flow shear. Recall that the large-$d_i$ limit of the homogeneous plasma dispersion relation is given by $\omega^2\approx k_\parallel^2\Omega_{ci}^2/k^2 + \mathcal{O}(1/d_i^4)$. Hence, as $d_i$ increases, the frequency should become independent of $d_i$. This is qualitatively consistent with the observed behavior in the HMHD eigenvalue calculations. Also, as $d_i$ increases, the range of $|B_\phi|$ corresponding to the unstable modes increases, and shifts to larger values. The change in $|B_\phi|$ partially compensates for the change in $d_i$ to produce approximately the same value of $\Omega_{ci}$.
\par
\begin{figure}[htbp]
    \centering
    \includegraphics[height=7cm]{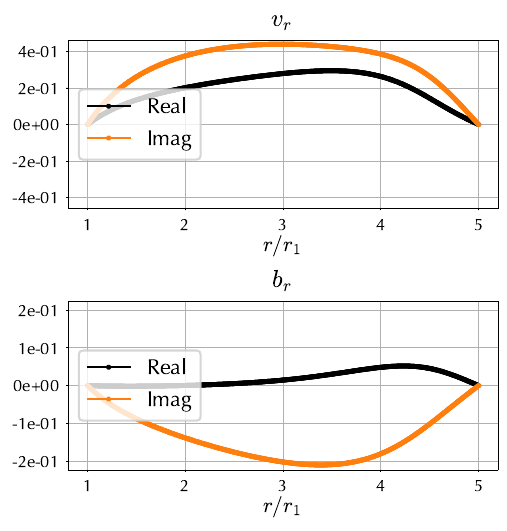}~\includegraphics[height=7cm]{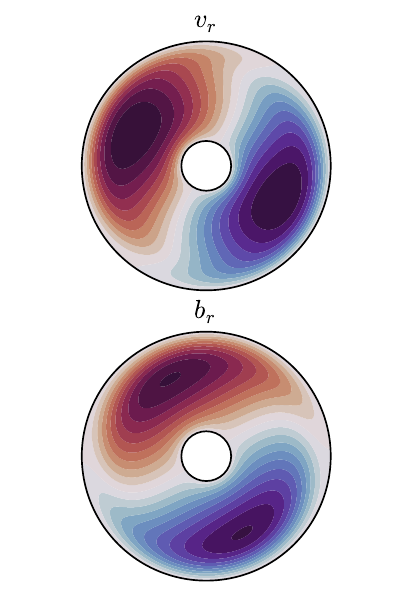}
    \caption{Radial profile (left) and contours in the $r-\phi$ plane (right) of the $m=1,k=\pi/4$ eigenfunction of the Hall-MHD system with $d_i=1r_1$ at $V_{A,\phi} = -2\Omega_0r_1$} 
    \label{fig: hmhd_azimuthal_di1.0_b2.0}
\end{figure}

 The polarization of the HMHD ion cyclotron wave in the large $d_i$ limit is such that the perturbed flow scales as $|\mb{v}|\sim |\bs{k}|d_i V_A$, and the magnetic field as $|\mb{b}|\sim |\bs{B}| $. Thus, using the normalized units, one expects that $|\bs{v}|/|\bs{b}| \sim kd_i V_A/r_1\Omega_0$. Figure \ref{fig: hmhd_azimuthal_di1.0_b2.0} plots the profiles of $v_r$ and $b_r$ from the most unstable mode at $V_{A,\phi}=-2\Omega_0r_1$ and $d_i=1$. The magnitude of $\bs{v}$ is approximately double the magnitude of $\bs{b}$, which is expected for the polarization of the ion-cyclotron for these parameters if $|\bs{k}| = \sqrt{m^2/r_1^2 + k^2}$ with $m=1,~k=\pm \pi/4$. Also plotted in figure \ref{fig: hmhd_azimuthal_di1.0_b2.0} are the resulting contour plots of the real parts of the perturbed fields after multiplying by $\exp(im\phi)$. 

\begin{deluxetable}{c|cc|cc}
\label{tab: nimrod}
    \tablecaption{Field strength and growth rates for linear NIMROD calculations in an azimuthal field. Each calculation has $d_i=1r_1$, $\beta=5\times10^6,\mathrm{Rm}\equiv r_1^2\Omega_0/\eta=2529$, $\mathrm{Pm}=1$. For each value of $V_{A,\phi}/r_1\Omega_0$, we compare to the results for the $k=\pm \pi/4$ modes from CYL\_SPEC}
\tablehead{
\colhead{$V_{A,\phi}/r_1\Omega_0$} \vline & \multicolumn{2}{c}{NIMROD (initial value)} \vline & \multicolumn{2}{c}{CYL\_SPEC (eigenvalue)}\\
 & \colhead{$\gamma/\Omega_0$} & \colhead{$|k|$ (observed)} \vline & \colhead{$\gamma/\Omega_0$} &\colhead{$k$ (input) }
}
\startdata
 -2.0 &   0.03 & $\pi/4$ & 0.03 & $\pi/4$ \\
 -0.5 &  0.05 &$\pi/4$ &  0.05 & $-\pi/4$ \\
 -0.1 &  0.13 & $\pi/2$ & 0.07 & $-\pi/4$ \\
  0.1 &  0.13 & $\pi/2$ & 0.07 & $\pi/4 $ \\
  0.5 &  0.05 & $\pi/4$ & 0.05 & $\pi/4$  \\
  2.0 &  0.03 & $\pi/4$ & 0.03 & $-\pi/4$   
\enddata
\end{deluxetable}

\par
\begin{figure}[htbp]
    \centering
    \includegraphics[width=0.3\linewidth]{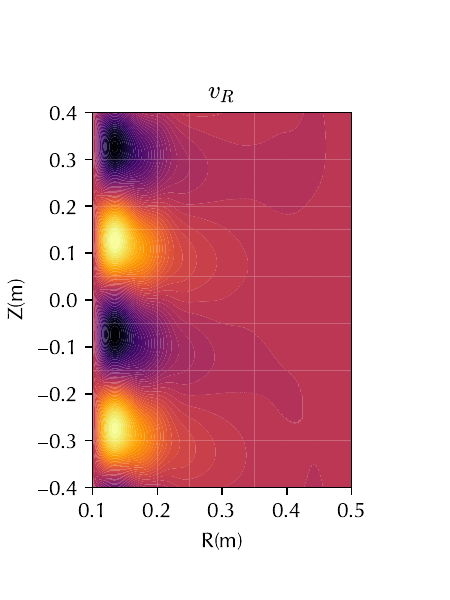}~\includegraphics[width=0.3\linewidth]{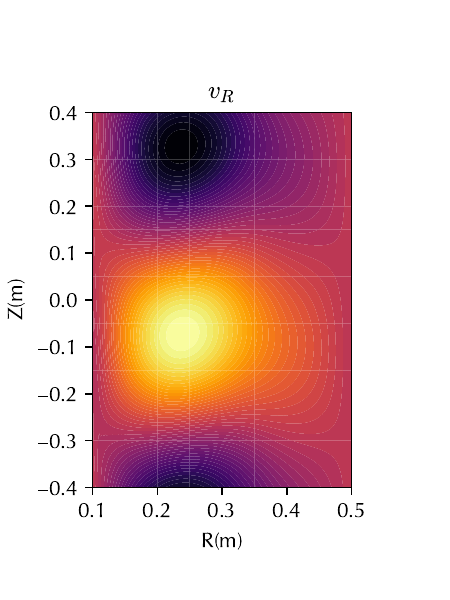}~\includegraphics[width=0.3\linewidth]{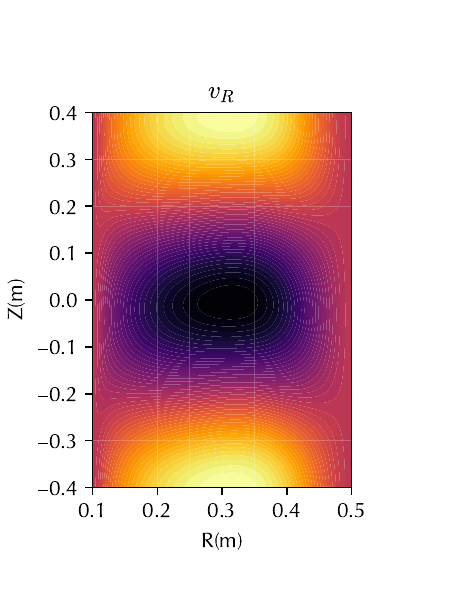}
    \caption{Contours of perturbed radial flow, $v_r$, in the $r-z$ plane from linear NIMROD calculations at $V_{A,\phi}/r_1\Omega_0=0.1, 0.5$, and $2$ (left to right). The lowest magnetic field strength case shows a faster growing mode with $k=\pi/2$, while the higher field cases show $m=1,k=-\pi/4$.}
    \label{fig: nimrod_rz_contours}
\end{figure}

\begin{figure}
    \centering
    \includegraphics[width=0.45\linewidth]{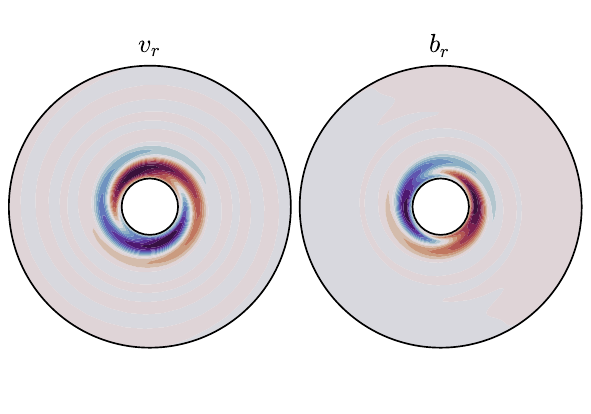}~\includegraphics[width=0.45\linewidth]{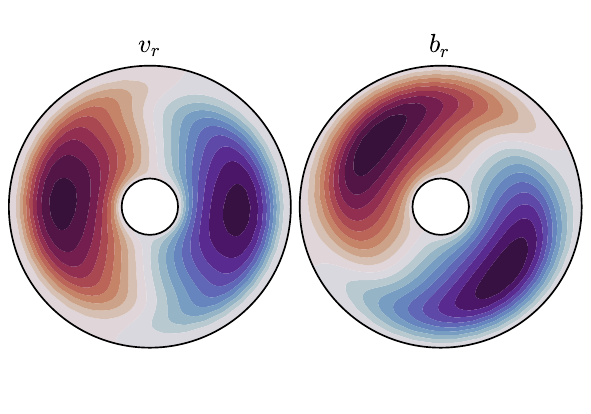}
    \caption{Contours of $v_r$ and $b_r$ in the $r-\phi$ plane from NIMROD calculations with $d_i=1.0$ when $V_{A,\phi}=0.1r_1\Omega$ (left) and $V_{A,\phi}=2r_1\Omega_0$ (right). The lower field case is more localized near the inner boundary. The higher field case agrees with the contours from CYL\_SPEC in figure \ref{fig: hmhd_azimuthal_di1.0_b2.0}}
    \label{fig: nim_azimuthal_di1_va2_eig}
\end{figure}

 Table \ref{tab: nimrod} summarizes the linear growth rates of the $m=1$ mode from NIMROD calculations in at a few values of azimuthal field strength for $d_i=1r_1$. The growth rates agree with the eigenvalue results at $V_{A,\phi}/r_1\Omega_0=\pm 0.5$ and $V_{A,\phi}/r_1\Omega_0=\pm 2$, but NIMROD reports higher growth for modes at $V_{A,\phi}/r_1\Omega_0=\pm 0.1$ magnetic field strength. This discrepancy is unsurprising, and owes to the fact that the axial dimension in NIMROD is discretized using finite elements. For a given azimuthal mode number, the calculated solution is a superposition of eigenmodes with several axial wavenumbers. The calculation of the growth rate is dominated by the axial wavenumber with the largest growth rate, which, for low magnetic fields, increases with $k$. This point is illustrated in figure \ref{fig: nimrod_rz_contours}, where we show contours of the radial component of $\bs{v}$ in the $r-z$ plane from NIMROD calculations at several field strengths. Visual inspection of the NIMROD solution at $V_{A,\phi}/r_1\Omega_0=0.1$ indicates that the dominant mode has axial wavenumber $k=2\pi/4$. Since we have only computed growth rates from CYL\_SPEC at $k=\pm\pi/4$, we should not expect agreement with the larger $k$ solution from NIMROD. Table \ref{tab: nimrod} also shows both the value of $|k|$ estimated by counting the number of vertical nodes in the perturbed NIMROD solution, and the input $k$ for the corresponding CYL\_SPEC calculation. 
At larger magnetic field strength, however, the NIMROD calculation finds that the fastest growing mode is the same $m=1,k=\pm\pi/4$ that was calculated using CYL\_SPEC. Figure \ref{fig: nim_azimuthal_di1_va2_eig} plots the radial component of the $m=1$ velocity and magnetic field perturbations from the NIMROD calculation with $V_{A,\phi}/r_1\Omega_0 = 0.5$ and $V_{A,\phi}/r_1\Omega_0 = 2$. 
The faster growing, high-$k$ mode observed when $V_{A,\phi}/r_1\Omega_0 = 1$ is radially localized near the point of maximum flow shear in the domain. At the larger field strength, the global mode is the most unstable.
~\\
\section{Electron-MHD limit -- unstable whistler modes}
\label{sec: emhd}
 In order to isolate the effect of whistler waves alone, and determine whether they couple to the flow shear to generate instability, we consider a limiting case of the Hall-MHD model. If the ion skin depth, $d_i$, is large compared to the wave-length corresponding to the dynamics of interest, say $\lambda$, then electron motion decouples from the ion motion, and we can neglect the terms containing $\mb{v}$ ( or $\bs{\xi}$ ) in the induction equation. This approximation yields the electron-MHD (EMHD) or EMH model~\citep{Gordeev1994}. This model can be considered valid in the range of plasma densities such that $d_i \gg \lambda \gg d_e$, or $m_ic^2/Ze^2 \gg n\lambda^2\gg m_ec^2/Ze^2 $. In terms of the parameters in equations \eqref{eq: hmhd-1}-\eqref{eq: hmhd-4}, it is appropriate to equate $\lambda$ with either the azimuthal or axial wavelength of the perturbations. Thus, the EMHD model should accurately describe perturbations with $d_im/r,~kd_i \gg 1$. In this limit, the system remains quasi-neutral, but the bulk ion motion is unimportant for the dynamics. Applying these assumptions to our problem of interest, we maintain the contribution from the inductive ${r\Omega}\phat\times\mb{b}$ in the Ohm's law, since it is the source of free energy driving instability. Neglecting $\mb{v}\times\mb{B}$ while maintaining ${r\Omega}\phat\times\mb{b}$ is consistent for modes where $|\omega|d_i \gg V_A $. It is shown in Section \ref{sec: global_ev} that for unstable for modes in this system, $|\omega| \sim |\Omega|$, so this approximation is valid at modest $d_i$ for weak magnetic fields, i.e. $V_A/r\Omega \ll 1$. Additionally, by neglecting the perturbed bulk plasma motion (i.e. the motion of the ion fluid,) the model describes only the coupling between the sheared, rotating flow and whistler waves. Applying these assumptions, equation \eqref{eq: basic-amp} becomes
\begin{equation}
    \partial_t\mb{b}+\Omega\partial_\phi\mb{b}-r\Omega'\phat b_r =  \frac{d_i}{\sqrt{\mu_0\rho}}\left(\curl{b}\cdot\nabla\mb{B}- 
      \mb{B}\cdot\nabla\curl{b}\right). \label{eq: emhd-amp}
\end{equation}
Equation \eqref{eq: emhd-amp}, together with $\nabla\cdot\bs{B}=0$, defines the EMHD approximation we will consider here.
\par In terms of the variables $\psi,\varphi,b_z$ used in Section \ref{sec: hmhd}, the radial and axial components of the equation \eqref{eq: emhd-amp} become
\begin{equation}
\obar \psi = -V_H(mb_z - k\varphi), \label{eq: emhd-r}
\end{equation}
\begin{equation}
	\obar rb_z = V_H\left(\varphi' - \frac{m}{r}\psi\right). \label{eq: emhd-z}
\end{equation}
The divergence constraint furnishes the third equation in lieu of the azimuthal component of the induction equation:
\begin{equation}
     \psi' - \frac{m}{r}\varphi - rkb_z = 0. \label{eq: emhd-div}
\end{equation}
The momentum evolution equation, \eqref{eq: basic-mom} does not appear here, since we neglect the effect of ion motion on the induction equation. From equations \eqref{eq: emhd-r}--\eqref{eq: emhd-div}, one can derive a second-order ODE for the variable $\psi \equiv irb_r$ given by 
    \begin{equation}
    	r|\mb{k}|^2\left(\frac{\psi'}{r|\mb{k}|^2}\right)' + \left(\frac{\obar^2}{V_H^2} + \frac{2mk\obar}{r^2|\mb{k}|^2V_H} - k\frac{r\Omega'}{V_H} + \frac{2kd_iB_\phi}{r\sqrt{\mu_0\rho}}\frac{\obar}{V_H^2} - |\mb{k}|^2\right)\psi = 0.
 \label{eq: emhd-psi}
    \end{equation}
 Here we have defined $|\mb{k}|^2\equiv m^2/r^2+k^2$. The terms in the parentheses of equation \ref{eq: emhd-psi} include both a term that depends on the flow shear, $r\Omega'$, and a term proportional to the curvature of the azimuthal magnetic field, $2B_\phi/r$. This equation is not immediately reducible to one describing a standard special function except when either $m=0$ or $k=0$ with particular choices for the $\Omega(r)$ profile. In particular, using the Keplerian rotation profile, $\Omega\propto r^{-3/2}$, solutions for $\psi$ are not immediately recognizable as standard special functions. Introducing the variable transformation $\psi = u(r)\exp(-1/2r|\mb{k}|^2 )$ transforms equation \eqref{eq: emhd-psi} to an equation for $u$ in which the coefficient of $u'$ vanishes. We also normalize the units using the inner wall dimension $r_1$ to normalize the radial coordinate and $1/\Omega_0 \equiv 1/\Omega(r_1)$ as a characteristic time scale. Following this procedure, we identify several dimensionless parameters that characterize the problem: $m$, $r_1k$, $\mathrm{H}\equiv V_H/r_1\Omega_0 = \omega_Ad_i/\Omega_0r_1$, $\Gamma(r) \equiv r\partial_r(\Omega/\Omega_0)$, and $ B_\phi/\sqrt{\mu_0\rho}\Omega_0r_1$. The dimensionless quantity $\mathrm{H}$ essentially measures the strength of the Hall EMF to the inductive EMF generated by flow shear in Ohm's law. The latter rotates the radial component of the perturbed magnetic field towards the azimuthal direction, while the former produces right-handed rotations of both components. It is the net effect of these two terms that results in overstable whistler modes.
 \par We will reuse $k$ and $\obar$ to mean the normalized parallel wave number and Doppler shifted frequency, respectively, i.e. $r_1k_1 \rightarrow k$, and $\obar/\Omega_0\rightarrow \obar$. In terms of the normalized quantities, equation \eqref{eq: emhd-psi} is transformed into
    \begin{equation}
    u'' + \left(\frac{\obar^2}{\mathrm{H}^2} + \frac{2mk\obar}{s^2|\mb{k}|^2\mathrm{H}} - \frac{k\Gamma}{\mathrm{H}}+\frac{2k\mathrm{H}_\phi}{s}\frac{\obar}{\mathrm{H}^2} - |\mb{k}|^2 + \frac{1}{4s^2} + k^2\frac{2m^2 - s^2k^2}{s^4|\mb{k}|^4}\right)u = 0, \label{eq: u-normalized}
\end{equation}
where $s\equiv r/r_1$ is the normalized radial coordinate, and $\mathrm{H}_\phi\equiv (B_\phi d_i/\sqrt{\mu_0\rho})/r_1^2\Omega_0$. Since we are dealing with axially periodic cylinders, $k = 2n\pi r_1/\ell $, for $n \in \mathbb{Z}$. So, for a given $r_1$ the parameter $k$ is determined by the axial mode number $n$, and the ratio $\ell/r_1$. The outer dimension of the cylinder $r_2$ also enters as a parameter via the boundary condition imposed at the outer wall. The aspect ratio, $A\equiv \ell/(r_2-r_1)$, is an additional dimensionless parameter on which the eigenvalues depend. Finally, note that although we have assumed that $d_i$ is large in deriving equation \eqref{eq: u-normalized}, the problem still depends on $d_i$ implicitly via $\mathrm{H}$. In principle changes in $|\mathrm{H}|$ could result from either a change in $|\bs{B}|$ or $d_i$. However, in order to maintain consistency with the large $d_i$ assumption, we must consider that $\mathrm{H}\rightarrow 0 \iff \bs{B}\rightarrow 0$.
\par Equations \eqref{eq: emhd-psi} and \eqref{eq: u-normalized} describe the global stability of non-axisymmetric whistler (HSI) instabilities in a differentially rotating cylindrical shear flow threaded with a current-free azimuthal, and uniform axial magnetic field. In the following, we will consider the local stability predictions of these equations followed by a discussion of global stability criteria and properties of unstable modes. Finally, we proceed with numerical eigenvalue calculations of the EMHD model using the CYL\_SPEC code.
\subsection{Local dispersion relation}
 Although it may sometimes lead to erroneous conclusions \citep{Knobloch1992a}, an approximate dispersion relation based on a WKB, wave-like ansatz is often used to extract stability information. In some instances the local stability analysis provides a necessary criterion for the existence of unstable modes. The EMHD model is amenable to this approach. Since we have already derived a single ODE that defines the eigenvalue problem, the error introduced by local approximation at this stage is less than would be incurred if the approximation were made earlier. Given a solution to equation \eqref{eq: u-normalized} that has locally oscillatory behavior, say $u''/u = -\delta^2 $, $\delta$ must satisfy the algebraic relation
\begin{equation}
    \frac{\obar^2}{\mathrm{H}^2} + \frac{2mk\obar}{s^2|\mb{k}|^2\mathrm{H}} - \frac{k\Gamma}{\mathrm{H}} +\frac{2k\mathrm{H}_\phi}{s}\frac{\obar}{\mathrm{H}^2} - |\mb{k}|^2-\delta^2 + \frac{1}{4s^2} + k^2\frac{2m^2-s^2k^2}{s^4|\mb{k}|^4} = 0.\label{eq: emhd-loc-quad}
\end{equation}
For a given $m,k,\delta,\Gamma$ and $\mathrm{H}$, this is a quadratic equation that determines $\obar$. Since the coefficients of equation \eqref{eq: emhd-loc-quad} are real, the sign of the discriminant determines whether there exist any unstable solutions ($\Im(\obar)>0)$. A necessary condition for complex roots is
  \begin{equation}
    -\frac{k\Gamma}{\mathrm{H}} > |\mb{k}|^2+\delta^2-\frac{1}{4s^2}-k^2\frac{2m^2-s^2k^2}{s^4|\mb{k}|^4} + \left(\frac{mk}{s^2|\mb{k}|^2}+\frac{k\mathrm{H}_\phi}{s\mathrm{H}}\right)^2 . \label{eq: emhd-local-crit}
\end{equation}
 Equation \eqref{eq: emhd-local-crit} implies that decreasing rotation profiles, $\Gamma <0$, with sufficiently strong shear are locally unstable. The first term on the right-hand side is simply the local frequency of a whistler wave. The other terms arise from the variable transformation from $\psi$ to $u$, and have the potential to be either stabilizing or destabilizing. The final two terms represent the effect of magnetic shear from the radially dependent azimuthal field. Note that for weak magnetic fields, $|\mathrm{H}| \rightarrow 0$, the instability criterion can be satisfied by modest shear.
Some aspects of this local analysis are qualitatively similar to the behavior of the local $m=0$ Hall-MRI dispersion relations~\citep{Wardle1999,Balbus2001}. For comparison, the $m=0$ limit of \eqref{eq: emhd-local-crit} is
\begin{gather}
    -\frac{r_1^2}{d_iV_{A,z}}\pdv{\Omega}{\ln s} >  |\mb{k}|^2 +\delta^2 + \frac{3}{4s^2}+\frac{B_\phi^2}{B_z^2}\frac{1}{s^2}. \label{eq: emhd-local-m0}
\end{gather}
Here we have used the simplification $\mathrm{H}\propto kd_iB_z $ when $m=0$. The fourth term on the right-hand side is essentially the rotational transform, or field-line twist, of the magnetic field, $\iota$. The expression in equation \eqref{eq: emhd-local-m0} is similar but not identical to the necessary criterion derived by \citet{Balbus2001} in the axisymmetric Hall-MHD case without the stabilizing contribution from magnetic field-line bending proportional to $(\mb{k}\cdot\mb{V}_A)^2$. It is also similar to the criterion given by Kunz~\citet{Kunz2008} for the HSI in a non-rotating sheared flow using the local shearing box approximation. The disappearance of the field-line bending term from equation \eqref{eq: emhd-local-m0} compared to equation (85) of \citet{Balbus2001} or equation (42) of \citet{Kunz2008} is a direct result of the EMHD approximation. The effect of magnetic shear is either stabilizing or destabilizing depending on the relative signs of $m$ and $\mathrm{H}_\phi/\mathrm{H}$. If the field is purely azimuthal, $\mathrm{H}_\phi/\mathrm{H}\propto 1/m$, and the contribution is  stabilizing. In the general case of $B_\phi\neq0,B_z\neq0$, there could exist a location where $m/s^2|\bs{k}|^2 + \mathrm{H}_\phi/s\mathrm{H} = 0 $, minimizing the stabilizing contribution of that term.
\par Finally, when equation \eqref{eq: emhd-local-crit} is satisfied, the solutions to equation \eqref{eq: emhd-loc-quad} are complex conjugates, and we can write the expression for the growth rate of the unstable solution as
\begin{equation}
    \gamma = |\mathrm{H}|\sqrt{\frac{2m^2-s^2k^2}{s^4|\mb{k}|^4}k^2-|\mb{k}|^2-\delta^2-\frac{k\Gamma}{\mathrm{H}}+\frac{1}{4s^2} -\left(\frac{mk}{s^2|\mb{k}|^2}+\frac{k\mathrm{H}_\phi}{s\mathrm{H}}\right)^2}. \label{eq: emhd-local-growth}
\end{equation}
We note that in the case of a purely axial magnetic field, this expression is does not depend on the sign of $m$ or $k$. In the purely azimuthal field case, however, the term $k\Gamma/\mathrm{H}=sk\Gamma/m\mathrm{H}_\phi$ depends on the relative signs of $m,k$ in relation to $\Gamma$ and $\mathrm{H}_\phi$.
\par Let us evaluate some limiting cases of equations \eqref{eq: emhd-local-crit} and \eqref{eq: emhd-local-growth}. As $k\rightarrow \infty $, we have $\mathrm{H}\sim kd_iV_{A,z}/r_1\Omega_0$, $\gamma \sim \sqrt{-k\Gamma\mathrm{H}-(k^2+(B_\phi^2/B_z^2)/s^2)\mathrm{H}^2}$. Thus, in general, large $k$ modes are only unstable for sufficiently small $|\mathrm{H}|$. Interestingly, nonzero $B_\phi$ will stabilize the high $k$ modes even as $B_z\rightarrow 0$ if $kd_iV_{A,\phi}/r_1\Omega_0>s\Gamma/4$. This stabilization of the high $k$ modes is independent of any dissipative mechanism (e.g. resistivity or viscosity). Also, if $|\mathrm{H}|$ is fixed, there are no growing modes for $k^2 > r_1\Omega_0\Gamma/d_iV_{A,z}$. When $m\rightarrow\infty$, equation \eqref{eq: emhd-local-crit} gives $-k\Gamma/\mathrm{H} > m^2/s^2 $, where $\mathrm{H}\rightarrow (m/s)d_iV_{A,\phi}/r_1^2\Omega_0$. The high-$m$ modes will certainly be stable for $|m| \gg |sk\Gamma r_1\Omega_0/ d_iV_{A,\phi} | $. For a given field strength, short wavelength whistler waves are stable.

\par The preceding local analysis of the EMHD model provides a dispersion relation the describes how three-dimensional whistler modes are destabilized by the flow shear, $\Gamma\equiv r\Omega'/\Omega_0$. By first reducing the linear system to a second-order differential equation for $\psi$, we are able to capture some non-local effects in the algebraic dispersion relation. Namely, the factors $1/4s^2$ and $k^2(2m^2-s^2k^2)/s^4|\bs{k}|^4$ in equation \eqref{eq: emhd-local-growth} are geometric effects from the curvilinear cylindrical coordinates that would not appear if we had assumed $\psi,\varphi$ and $b_z$ had exponential dependence on $s$ at the level of equations \eqref{eq: emhd-r}--\eqref{eq: emhd-psi}. 
\par The local criterion for instability in equation \eqref{eq: emhd-local-crit} and subsequent expression for the growth rate (equation \eqref{eq: emhd-local-growth}) provide physical intuition, but extracting quantitative results from these expressions presents some difficulty. There is an ambiguity both in the choice of the radial wavenumber, $\delta$, as well as the radial location at which the various $s$-dependent terms should be evaluated. Moreover, we observe in numerical calculations that unstable eigenmodes can extend throughout the domain -- invalidating assumptions of locality. \textit{It is also important to note that a failure to meet the local instability criterion in equation \eqref{eq: emhd-local-crit} does not necessarily preclude the existence of global unstable eigenmodes.} In particular, when $k=0, m\neq0$ in the azimuthal magnetic field case, there are unstable global modes associated with whistler waves being amplified by the flow. This specific example is discussed further at the end of Section \ref{sec: emhd_k0}.

\subsection{Global Eigenvalue Problem}
\label{sec: global_ev}
 In this section we develop some analytic properties of the EMHD eigenvalue problem that hold for any eigenmode, and show examples of unstable eigenmodes obtained from numerical solutions of equation \eqref{eq: emhd-psi}
\par General conclusions about any potentially unstable modes can be extracted by multiplying either eqn. \eqref{eq: emhd-psi} or equation \eqref{eq: u-normalized} by $\psi^*$ or $u^*$ respectively, integrating over the domain, and applying the boundary conditions $\psi=0$ at $r=r_1,r=r_2$ or $u=0$ at $s=1$ and $s=r_2/r_1$. Applying this procedure to equation \eqref{eq: u-normalized}, and taking the imaginary part, we find 
\begin{equation}
    \gamma\int_{1}^{r_2/r_1} \left( \omega_r - m\Omega + \frac{mk\mathrm{H}}{s^2|\mb{k}|^2}+\frac{k\mathrm{H}_\phi}{s}\right)\frac{|u|^2}{\mathrm{H}^2} ds = 0. \label{eq: emhd-stab-criterion}
\end{equation}
For $\gamma \neq 0$, this expression constrains the relation between the rotation profile, the whistler phase velocity, and the real frequency of an unstable mode. Since $|u|^2>0$, the expression in the parentheses must change sign somewhere in the domain. This is a necessary condition, and its utility is limited since it depends on both the real part of the unknown eigenvalue, and the radial profile of the unknown eigenfunction. Equation \eqref{eq: emhd-stab-criterion} does, for a given $\Omega, m, k,$ and $\mathrm{H}$, provide a bound on the range of $\omega_r$ over which the whistler modes can possibly be unstable.
\par A similar procedure can be used to derive a quadratic equation for $\omega$ with coefficients that are integrals over the domain that involve the unknown function $u$. Specifically, it can be shown that $\omega \in \mathbb{C}$ satisfies
\begin{equation}
    \mathcal{A}\omega^2 + \mathcal{B}\omega+ \mathcal{C} = 0, \label{eq: emhd-quad}
\end{equation}
where
\begin{displaymath}
    \mathcal{A}\equiv \int \frac{|u|^2}{\mathrm{H}^2}~ds ,\;\;
    \mathcal{B}\equiv 2\int \left(\frac{mk\mathrm{H}}{s^2|\mb{k}|^2}+\frac{k\mathrm{H}_\phi}{s}-m\Omega\right)\frac{|u|^2}{\mathrm{H}^2} ~ds ,
\end{displaymath}
    and
    \begin{equation}
    \mathcal{C} \equiv \int \left(m^2\Omega^2-2m\Omega\left(\frac{mk\mathrm{H}}{s^2|\mb{k}|^2}+\frac{k\mathrm{H}_\phi}{s}\right) -k\Gamma\mathrm{H} -\mathrm{H}^2\left(|\mb{k}|^2-\frac{1}{4s^2} - k^2\frac{2m^2 - s^2k^2}{s^4|\mb{k}|^4}\right)\right)\frac{|u|^2}{\mathrm{H}^2}~ds -\int |u'|^2 ~ds.
    \label{eq: emhd-ccoef}
\end{equation}
Since $\mathcal{A,B,C}\in\mathbb{R}$, we can conclude that a necessary and sufficient condition for stability is that $\mathcal{B}^2 -4\mathcal{AC} \geq 0$. Moreover, since $\mathcal{A}>0$ a sufficient condition for stability is $\mathcal{C}\leq0$. Unlike equation \eqref{eq: emhd-stab-criterion}, evaluating these conditions does not require the value of $\Re{\omega}$, but it does rely on knowledge of $|u|^2$ and $|u'|^2$. Since the term involving $\int |u'|^2$ appears with minus sign in the expression for $\mathcal{C}$, it represents a stabilizing effect on localized or highly oscillatory solutions. If, however, the solution is localized in a region of sufficiently strong flow shear, the stabilizing contribution from $-\int |u'|^2$ can be counteracted by $\int -k\Gamma |u|^2/\mathrm{H}$. Also, note that eqs. \eqref{eq: emhd-quad}-\eqref{eq: emhd-ccoef} assume the boundary conditions $u(1)=u(r_2/r_1)=0$. Different choices of boundary condition will yield different stability criteria. 

\begin{figure}[htbp]
    \centering
    \includegraphics[width=0.8\linewidth]{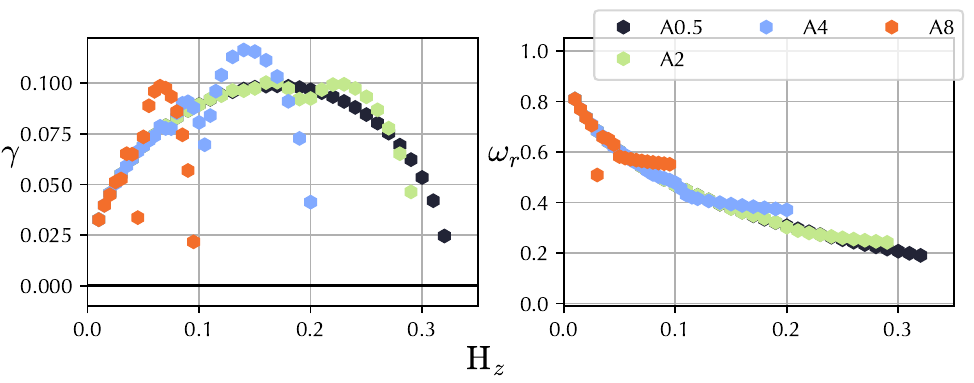}
    \caption{Numerically computed growth rates and frequencies of the fastest growing radial mode in a disk with axial magnetic field in the electron MHD limit with $m=1,k=\pi/4 $ as a function of $\mathrm{H}\propto kB_z$ at several aspect ratios.}
    \label{fig: emhd_axial_m1k1_ascan}
\end{figure}
\subsection{Numerical Results for EMHD}
\label{sec: emhd_numeric}
\par   We present numerical results from the electron MHD system using another version of the CYL\_SPEC code described in Section \ref{sec: hmhd-numeric} 
\par
\begin{figure}[hbt!]
    \centering
    \includegraphics[width=0.4\linewidth]{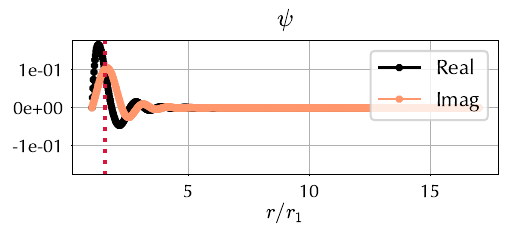}\;~\;\includegraphics[width=0.4\linewidth]{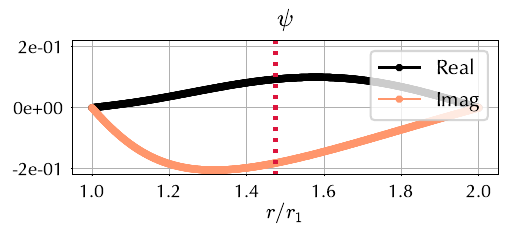}
    \caption{Plots of the eigenfunction $\psi$ associated with the fastest growing modes from the axial field case at $A=0.5$, $\mathrm{H}=0.08$ (left), and $A=8$, $\mathrm{H}=0.08$ (right). The dashed red lines denote the point $\obar=0$ for each mode. Note the difference in radial scales. }
    \label{fig: emhd_eig_compare}
\end{figure}

 Here, we consider the most global non-axisymmetric mode, i.e. $m=1,k=2\pi r_1/\ell$. This value of $k$ corresponds to the minimum non-zero normalized axial wavenumber that satisfies periodic boundary conditions at the ends of the cylinder. In figure \ref{fig: emhd_axial_m1k1_ascan}, we plot the growth rate of the fastest growing numerically computed mode as a function of $\mathrm{H}$ at several values of the aspect ratio, $A$, defined as $A\equiv \ell/(r_2-r_1)$. The variation in aspect ratio is achieved by fixing $\ell/r_1=8 \implies k=\pi/4$ and adjusting the distance between the inner and outer boundary, $r_2/r_1$.
As the aspect ratio (i.e. disk thickness) increases, the outer boundary is moved closer to the inner boundary, and the range of $\mathrm{H}$ that supports unstable modes decreases. In the thin disk limit limit $A\ll 1$, the outer boundary is far enough away that it does not affect the fastest growing mode.
Figure \ref{fig: emhd_eig_compare} compares plots of the numerically computed radial eigenfunction $\psi$ of equation \eqref{eq: emhd-psi} a low aspect ratio ($A=0.5$) and a high aspect ratio ($A=8$) case. In the high aspect ratio case, the unstable mode is localized near the inner boundary, and is substantially damped to the right of the co-rotation point. In the low aspect ratio case, the co-rotation point is closer to the outer boundary, and the mode has nontrivial amplitude throughout the domain. 
\par
\begin{figure}
    \centering
    \includegraphics[width=0.8\linewidth]{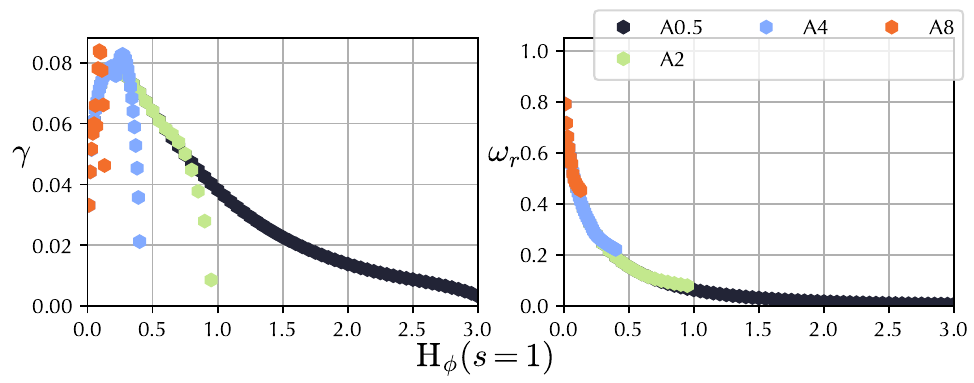}
    \caption{Numerically computed growth rates and frequencies of the fastest growing radial mode in a disk with azimuthal magnetic field in the electron MHD limit with $m=1,k=\pi/4$ as a function of $\mathrm{H}\propto B_\phi$ evaluated at $s=1$ at several aspect ratios.}
    \label{fig: emhd_azimuthal_m1k1_ascan}
\end{figure}

 Turning to the disk configuration with purely azimuthal magnetic field, we note two unique feature that are absent in the uniform axial field case. Since $\mb{B} \parallel \phat $, the propagation direction of whistler waves is also along the direction of the bulk flow. Also, the magnetic field strength varies with radial location in accordance with the current-free condition $(rB_\phi)' = 0$. In this case, there is also a transition from the low-field, local modes to higher field global modes that depends on the aspect ratio. Figure \ref{fig: emhd_azimuthal_m1k1_ascan} plots the growth rates and frequencies of the $m=1,k=\pi/4$ mode
 Since the location of the outer boundary does not appear explicitly in in eqs. \eqref{eq: emhd-local-crit} or \eqref{eq: emhd-local-growth}, one would not expect to be able to explain this behavior based on local theory alone. Qualitatively, one can infer some effect by introducing a heuristic expression for the radial wavenumber $\delta \propto 1/(r_2/r_1-1)$ into the local dispersion relation, equation \eqref{eq: emhd-local-growth}. With this expression for $\delta$, decreasing aspect ratio yields a smaller $\delta^2$. For a fixed azimuthal mode number, $m$, the minimum value of $m^2/r^2$ attained in the domain also decreases with decreasing aspect ratio. These two effects result in a correspondingly smaller value for $|\mb{k}|$ appearing on the right-hand side of equation \eqref{eq: emhd-local-growth}, suggesting larger local growth rates.

 \subsection{Over-reflection of \texorpdfstring{$k=0$}{k=0} modes in azimuthal field} 
 \label{sec: emhd_k0}
 When $B_\phi\neq 0$, the EMHD model has unstable solutions that correspond to unstable modes with $k=0$ and $m\neq0$. Instability in this case is an interesting finding for two reasons. First, since the term that contains the flow shear, $k\Gamma/\mathrm{H}$, vanishes when $k=0$, one may be under the impression that the instability drive is absent. It is, however, not explicitly required that $k\Gamma/\mathrm{H}\neq0$ for instability to occur. Second, when $k=0$ the local criterion in equation \eqref{eq: emhd-local-crit} predicts stability for $m^2+s^2\delta^2 > 1/4$. Since eigenmodes in the cylindrical domain are $2\pi$-periodic in $\phi$, the  minimum nonzero value of $m^2$ permitted is $m^2=1$. Therefore, any choice of $\delta \in \mathbb{R}$ is stable. The existence of unstable modes with $m=1,k=0$ highlights the potential danger of drawing stability conclusions based solely on local WKB analysis.
 \par Using the global criteria based of equation \eqref{eq: emhd-stab-criterion}, we find that $\gamma \neq 0$ is possible if $\int_1^{r_2/r_1} (\omega_r-m\Omega(s))s^4|u|^2 ds =0$. The important condition in this case is whether the rotation profile varies sufficiently so that $\omega_r - m\Omega(r)$ changes signs in the domain. This requirement for instability suggests that the unstable modes are whistler waves being destabilized by a `corotation amplifier' mechanism~\citep{Narayan1987,Tsang2008}. 
\begin{figure}[htbp]
     \centering
     \includegraphics[width=0.45\linewidth]{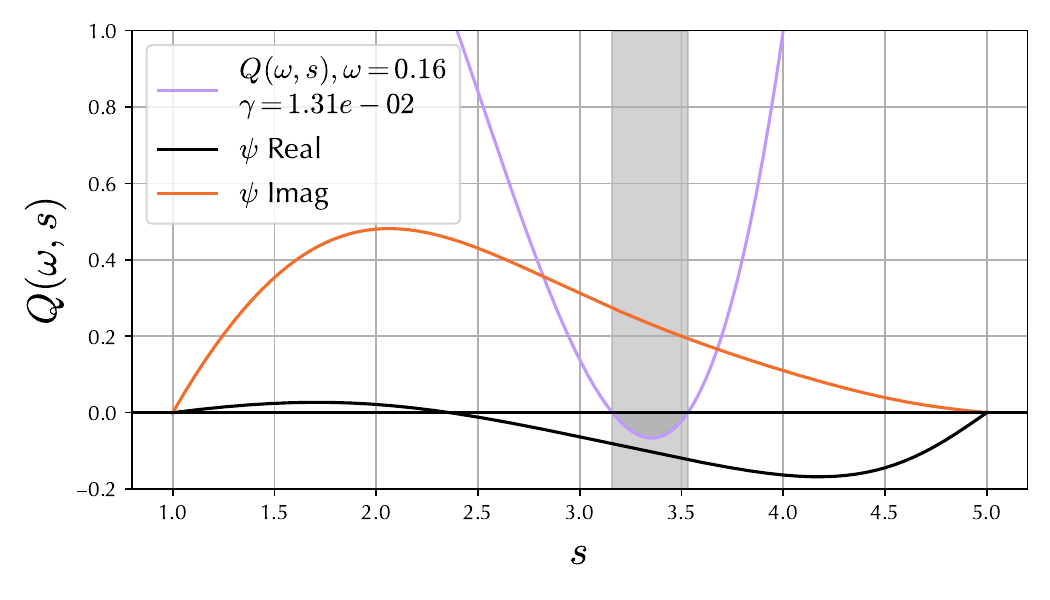}~\includegraphics[width=0.45\linewidth]{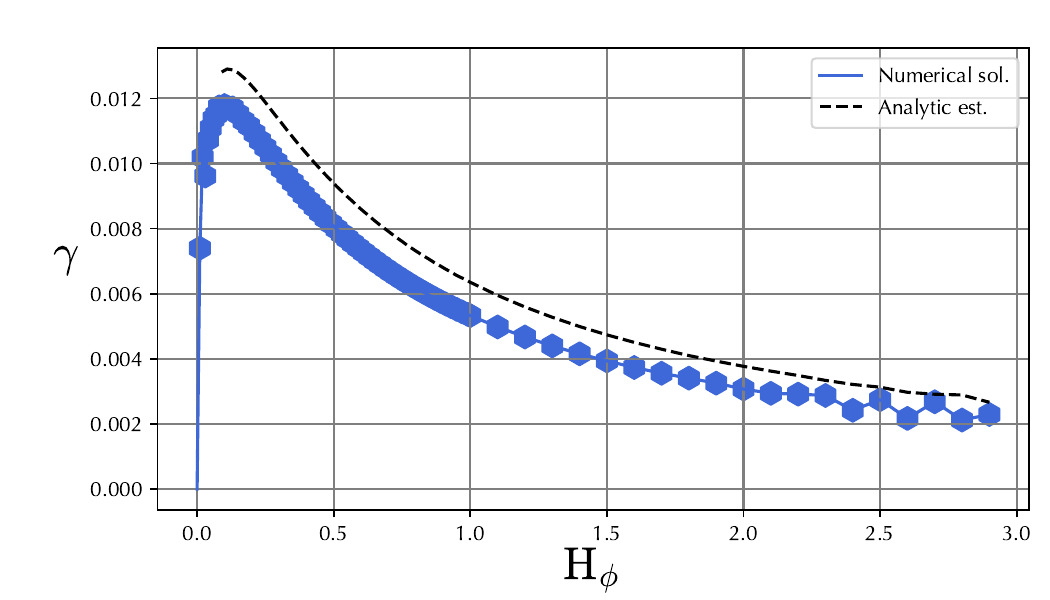}
     \caption{Left: Plot of the square of the effective refractive index for whistler waves, $Q(\omega,s)$ for the fastest growing $m=1,k=0$ mode with $d_iV_{A,\phi}=0.58$. The shaded region is where $Q(\omega,s)<0$, and the zeros of $Q$ are where $\obar^2 = |\bs{k}|^2d_i^2\omega_A^2$ The real and imaginary parts of $\psi$ are also plotted. Right: plot of the growth rate for $m=1,k=0$ modes in the EMHD model as a function of $\mathrm{H}_\phi$ for aspect ration $A=0.5$ along with analytic estimate based on equation \eqref{eq: gamma_one_wall}.}
     \label{fig: emhd-k0-potential-plot}
 \end{figure}
 For modes of this type, the term multiplying $u$ in equation \eqref{eq: u-normalized}, which is the square of the refractive index for whistler waves, can become negative. Denote this term by the function $Q(\omega,s)$. Then, in regions where $Q<0$, the radial wave-vector becomes imaginary for real $\omega$, and the whistler modes suffer spatial damping.
 
 Figure \ref{fig: emhd-k0-potential-plot} plots the function $Q(\omega,s)$ corresponding to the fastest growing mode attained by this system. Also plotted is the structure of eigenmode. Note that the `potential well' in this case is relatively shallow. The phase of the eigenmode fortuitously illustrates the transition between the imaginary being dominant in the region left of the first turning point, and the real part dominant to the right of the second turning point. This is what one qualitatively expects by applying asymptotic (WKB) turning point theory. As shown in appendix \ref{sec: whistler-app}, the solutions may be approximated by parabolic cylinder functions that are oscillatory in the region outside of the shaded area, and behave as growing and decaying exponential functions inside the shaded region. Extending the discussion of wave over-reflection generating global modes in non-conducting fluid disk model to the present EMHD whistler model in a manner similar to \citet{Narayan1987} is conceptually straightforward. The technical details and analysis of equation \eqref{eq: u-normalized} are discussed in appendix \ref{sec: whistler-app}.

\section{Conclusion}
We have investigated the stability of rotating plasmas in regimes beyond MHD using analytical and numerical analysis via both eigenvalue and initial value calculations. We find that Hall-MHD leads to additional global non-axisymmetric instabilities in a differentially rotating plasma. The Hall EMF breaks the degeneracy of the linear incompressible MHD dispersion relation, yielding two separate branches that, in the large $|\bs{k}|d_i$ limit, become whistler and ion-cyclotron waves. Both of these waves can be destabilized by the flow shear. The coupling of the Hall-MHD waves to the background flow shear leads to new branches of global instabilities. In both magnetic geometries -- axial and azimuthal -- increasing the magnetic field strength preferentially stabilizes small wavelength modes, allowing the large-scale, global modes to dominate. We make two interesting observations about these global, non-axisymmetric Hall-MHD modes. First, there are some parameters at which they grow much faster than non-axisymmetric modes in the MHD model. The second is that these modes may be unstable at much larger field strengths than any instabilities in the MHD model.

In the full Hall-MHD system, we varied the strength of the hall term by adjusting $d_i/r_1$. When $d_i/r_1 = 1$, the fastest growing non-axisymmetric modes are unstable whistler waves at weak magnetic field. This is confirmed through comparison with the results from the EMHD model. At higher field strengths, the whistler modes are stabilized, and global, long-wavelength `ion-cyclotron' modes are the most unstable.
As the ion skin depth is increased, the system  supports non-axisymmetric instabilities at increasingly larger magnetic field strengths. The range of ion-cyclotron instability in the azimuthal field case, in particular, moves towards values of $|B_\phi|$ corresponding to sub-Alfv\'enic flow. This general observation is consistent with the notion that the Hall effect decouples the ion motion from the magnetic field, so that the stabilizing contribution from field-line bending is weaker than in the MHD model. The numerical results from CYL\_SPEC have been verified by comparison with initial value calculations using the NIMROD code, where the same growth rates and similar radial mode structures where obtained. 

By appealing to the EMHD limit, where the dynamical coupling between the electrons and the ions is weak, we studied the subset of HMHD modes corresponding to whistler waves that are driven unstable by the flow shear. The local analysis and stability criteria derived from the EMHD are a generalization of previous discussions of the Hall-shear or magneto-shear instabilities~\citep{Kunz2008,Bejarano2011} to the non-axisymmetric case in a rotating flow, including the effect of geometric curvature. The numerical calculations interrogated the dependence of the whistler modes on the aspect ratio of the disk geometry, finding that lower aspect ratios support growing modes at larger field strengths. Both local and global stability criteria for the EMHD system are presented.
\par A subset of the unstable whistler modes with $k=0$ in the azimuthal field case are shown to be destabilized by a `corotation amplifier' mechanism similar to those discussed by \citet{Lindzen1978}, \citet{Narayan1987}, and \citet{Tsang2008}. Crucially, unstable modes of this type are not predicted by local dispersion relations, and have not been previously identified.

\par The Hall term modifies the linear mode spectrum of non-axisymmetric modes at relatively modest values of the normalized ion skin-depth. We have normalized $d_i$ in terms of the radius of the inner wall of the annulus, but it is more common to discuss how it compares to the disk scale height. In our EMHD and Hall-MHD calculations, the height of the domain is $\ell = 8r_1$. The values of $d_i/\ell $ considered are $1.25\times10^{-3}, 1.25\times10^{-2}$ and $1.25\times 10^{-1}$. We compare this to estimates for protoplanetary disks used by \citet{Kunz2013}, where the relevant parameter is $\ell_H\equiv \sqrt{(\rho/\rho_i)} d_i $. Based on a minimum-mass solar nebula, \citet{Bthune2016} gives an estimate of the range of $\ell_H/h$, where $h$ is the scale height of the disk, between $10^{-2}$ and $10^2$. Thus, the values of $d_i$ and the corresponding low-frequency global Hall-MHD modes obtained here may be relevant to protoplanetary disks.

\par Subsequent work will address how the more virulent non-axisymmetric linear instabilities in the Hall-MHD system may affect nonlinear momentum transport. Global nonlinear Hall-MHD NIMROD calculations are ongoing that aim to quantify the level of momentum transport that can be attributed to the global, non-axisymmetric modes. Concerning linear theory -- since we are interested in global stability of the differentially rotating system, we aim to extend these results by considering how they are affected by inhomogeneities in the steady-state (radial and axial stratification, gradients in the magnetic field, etc.). 

\begin{acknowledgments}
This work is funded by the U.S. National Science Foundation award No. 2308829.\\

\noindent This work used the high-performance computing cluster at Princeton Plasma Physics Laboratory \\
    
\noindent This work used Bridges-2 at Pittsburgh Supercomputing Center through allocation phy240045p from the Advanced Cyberinfrastructure Coordination Ecosystem: Services \& Support (ACCESS) program, which is supported by National Science Foundation grant Nos. 2138259, 2138286, 2138307, 2137603, and 2138296.
\end{acknowledgments}
\appendix 
\section{Whistler wave over-reflection in the EMHD model}
\label{sec: whistler-app}
When $k=0,B_z=0$, but $ m\neq0, B_\phi\neq0$, equation \eqref{eq: u-normalized} becomes
\begin{equation}
   u'' + \left( \frac{\obar^2}{m^2V_{A}^2}\frac{r_1^2}{d_i^2}s^4 - \frac{m^2-1/4}{s^2}\right) u = 0. \label{eq: k=0}
\end{equation}
Here we have assumed the form $B_\phi = B_\phi(r_1)r_1/r$, and set $V_A \equiv B_\phi(r_1)/\sqrt{\muo\rho}/r_1\Omega_0$. Define the effective refractive index for whistler modes, $Q(\omega,s)\equiv \frac{\obar^2}{m^2V_{A}^2}\frac{r_1^2}{d_i^2}s^4 - \frac{m^2-1/4}{s^2} $. For the Keplerian rotation profile $\Omega/\Omega_0 = s^{-3/2}$, and real $\omega$, the function $Q$ can have either three zeroes, two zeros, or one zero in the domain $1\leq s\leq r_2/r_1$. Note that the zeros of $Q$ are points where the local Doppler-shifted frequency is equal to the whistler frequency, $\obar^2 = \omega_H^2 \equiv |\bs{k}|^2 d_i^2k_\parallel^2 V_A^2$  Given $\omega$, there exists a sufficiently large $V_A$, beyond which only one root is possible. The value of $V_A$ at which this transition occurs can be computed by analyzing the equation
\begin{equation}
    \omega^2s^6 - 2m\omega s^{9/2} + m^2s^{3} - \frac{m^2d_i^2V_A^2}{r_1^2}\left(m^2-\frac{1}{4}\right) = 0.
\end{equation}
This can be rewritten as a bi-quadratic equation in the variable $\xi = s^{3/2}-m/2\omega$.
\begin{equation}
    \xi^4 - \frac{m^2}{2\omega^2}\xi^2 + \frac{m^2}{\omega^2}\left(\frac{m^2}{16\omega^2} - \frac{d_i^2V_A^2}{r_1^2}\left(m^2-\frac{1}{4}\right)\right) = 0 \label{eq: biquad}
\end{equation}
In order to have two roots in the domain $1< s < r_2/r_1 $, we need real positive roots $\xi$ in the range $1-m/2\omega \leq \xi \leq (r_2/r_1)^{3/2} - m/2\omega $. We know from equation \eqref{eq: emhd-stab-criterion} that in this case we require $\obar =0$ somewhere in the domain in order for an unstable mode to exist. So for unstable modes $\omega$ is bound by $m \leq \omega \leq m (r_1/r_2)^{3/2}$. Thus, we are concerned with the number of real roots, $\xi$, in the range $1/2 \leq \xi \leq (r_2/r_1)^{3/2} /2$. Based on D\'escartes' rule of signs, we know that there is exactly one real root $\xi^2$ in equation \eqref{eq: biquad} if 
\begin{gather}
    \frac{m^2}{16\omega^2} - \frac{d_i^2V_A^2}{r_1^2}\left(m^2-\frac{1}{4}\right) < 0.
\end{gather}
This means that for
\begin{gather}
    (d_i/r_1)^2V_A^2 > \frac{(r_2/r_1)^3}{16}\frac{m^2}{m^2-1/4},
\end{gather}
the function $Q$ has at most one zero in the domain. When $V_A\ll 1$, there are generally two positive roots $\xi$, giving two roots for $s$, and one negative root $\xi$ such that $\xi + m/2\omega > 0$, which gives another valid root for $s$. The most common case has the third root appearing outside the domain at some $s<1$. For a given $V_A$, one could alternatively consider the ranges $\omega$ for which there are one, two or three zeros.
\par Having investigated the structure of the refractive index $Q$, the problem can be treated using WKB theory. The resulting estimates of eigenvalues will depend on the number of zeros of $Q$ present in the domain. It should be noted that we although we have assumed real $\omega$ the following analysis is valid when $\Im{\omega}\ll \Re{\omega}$. If the growth rate or damping rate is equal to or exceeds the frequency of the mode, the turning points move away sufficiently far from the real axis, and the analysis must be modified. We consider the one and two turning point cases. The three turning point case is more complicated, and does not appear necessary to describe the unstable modes found numerically in Section \ref{sec: emhd_k0}.
\subsection{One turning point case}
In the case there is only one zero of $Q$ in the domain, there is a single turning point. Denote the zero of $Q$ by $\mu$. Then, in this case, the solutions are exponential in the region $1\leq s \leq \mu$, and oscillatory in the region $s_0\leq s\leq r_2/r_1$. Using the technique of \citet{Langer1959} we can approximate the solution in this case as 
\begin{equation}
    u \sim \frac{1}{\sqrt{\zeta'(s)}}\left(c_1 \mathrm{Ai}(-\zeta(s) ) + c_2\mathrm{Bi}(-\zeta(s)) \right).
\end{equation}
where $\mathrm{Ai},\mathrm{Bi}$ are the usual Airy functions, and $\zeta$ is defined by
\begin{equation*}
    \zeta^{3/2} = \frac{3}{2}\int_{\mu}^s \sqrt{Q(\omega,\xi)}~d\xi, \;\;\text{for}\;\; s \geq \mu,
\end{equation*}
and
\begin{equation*}
    (-\zeta)^{3/2} = \frac{3}{2}\int_{s}^{\mu} \sqrt{-Q(\omega,\xi)}~d\xi, \;\;\text{for}\;\; s \leq \mu.
\end{equation*}
The signs have been chosen so that the solution is oscillatory in the region region $s\geq \mu$. Enforcing the boundary conditions at $s=1$ and $s=r_2/r_1$ gives the requirement
\begin{equation}
    \frac{\mathrm{Ai}(-\zeta(r_2/r_1))}{\mathrm{Bi}(-\zeta(r_2/r_1))} = \frac{\mathrm{Ai}(-\zeta(1)) }{\mathrm{Bi}(-\zeta(1))}.
\end{equation}
There are two interesting limiting cases depending on the location of $\mu$ in the domain. The first is when the turning point is located closer to the outer boundary -- $\mu-1 \ll r_2/r_1 - \mu$. Then $\mathrm{Ai}(-\xi(1))/\mathrm{Bi}(-\xi(1))\rightarrow 0$. Then $-\zeta(r_2/r_1)$ must be a zero of the Airy function $\mathrm{Ai}$. Since the zeros of $\mathrm{Ai}$ are all real, we have that
\begin{equation*}
    \int_\mu^{r_2/r_1} \sqrt{Q(\omega,\xi)}~d\xi \in \mathbb{R}.
\end{equation*}
If $\Im{\omega}=0$, then $\Im{Q(\omega,\xi)}=0$, and this can be satisfied. If $\Im{\omega}\neq 0$ the only way to satisfy the dispersion relation is to have $\int \Im{\sqrt{Q}} = 0$, which requires $\Im{Q}$ to change sign at some point in the range $\mu\leq s\leq r_2/r_1$. Since $\Im{Q} \propto (\omega_r-\Omega)$, $\Im{Q}$ cannot change sign in the region $s\geq \mu $. A similar argument holds when the turning point is situated close to the inner boundary.
\par If the location of the turning point is sufficiently far from both boundaries, the asymptotic behavior of the Airy functions gives the relation 
\begin{equation}
    \tan(\int_\mu^{r_2/r_1} \sqrt{Q}~d\xi -\frac{\pi}{4}) = -2\exp(2\int_1^\mu \sqrt{-Q}~d\xi ),
\end{equation}
or, writing $\arctan $ in terms of a complex-valued logarithm, we have
\begin{equation}
    \int_\mu^{r_2/r_1} \sqrt{Q}~d\xi -\frac{\pi}{4} =\frac{1}{2i}\log(\frac{1-2i\exp(2\int_1^\mu \sqrt{-Q}~d\xi )}{1+2i\exp(2\int_1^\mu \sqrt{-Q}~d\xi )}).
\end{equation}

\subsection{Two turning point case}
When $V_A \ll 1$, there are two roots of $Q$ located at $\left(\frac{m}{2\omega}\left( 1 + \sqrt{1 \pm 4\omega d_iV_A\sqrt{m^2-1/4}} \right)\right)^{2/3}$. In this case, the domain is divided into two regions where the solution is wavelike, $Q>0$ separated by an `evanescent region' region where the solution is damped, $Q<0$. Let $\mu_1,\mu_2$ denote the locations of the two zeroes of $Q$, and assume $Q<0$ for $\mu_1\leq s\leq \mu_2$. Then, it can be shown using a Liouville-Green transformation that the solutions are asymptotic to parabolic cylinder functions ~\citep{Nayfeh2000}. In particular, define $\zeta(s)$ such that when $\mu_1 \leq s\leq \mu_2$
\begin{equation*}
    2a\int_{-1}^{\zeta} \sqrt{1-\xi^2}~d\xi = \int_{\mu_1}^s\sqrt{-Q(\omega,\xi)} ~d\xi,
\end{equation*}
where $a$ is a constant chosen to so that the interval $\mu_1\leq s\leq \mu_2$ is mapped to $-1\leq \zeta \leq 1$. The required value of $a$ is
\begin{equation*}
    a \equiv \frac{1}{\pi}\int_{\mu_1}^{\mu_2} \sqrt{-Q}~d\xi.
\end{equation*}
$a$ can be interpreted as the size of the `classically forbidden region' in a quantum mechanical tunneling problem. Similar expressions involving suitably chosen branches of the square roots ensure $\zeta(s)$ is a regular function, and that the regions where $Q>0$ are mapped to $\zeta^2>1$, and the region where $Q<0$ is mapped to the region $\zeta^2 < 1$. Namely, for $1\leq s\leq \mu_1$, set
\begin{equation*}
    2a \int_{1}^\zeta \sqrt{\xi^2-1}~d\xi =  \int_{\mu_1}^s \sqrt{Q}~d\xi.
\end{equation*}
And when $\mu_2 \leq s \leq r_2/r_1$ set
\begin{equation*}
    2a \int_{1}^\zeta \sqrt{\xi^2-1}~d\xi =  \int_{\mu_2}^s \sqrt{Q}~d\xi.
\end{equation*}
With this variable transformation, $\zeta$ is a smooth function of $s$ and the original equation for $u$ can be transformed into the following related equation for $v(\zeta)$, where $v \equiv (4a^2(\zeta^2-1)/Q)^{1/4}u$:
\begin{equation}
    v'' + 4a^2(\zeta^2-1)v = 0,
\end{equation}
where , is the normalization factor required to map the region $Q<0$ to the interval $-1\leq \zeta \leq 1$.
The solutions $v$ are parabolic cylinder functions 
\begin{equation*}
    v = c_1E\left(a, 2\sqrt{a}\zeta\right)+ c_2E^*\left(a,2\sqrt{a}\zeta\right),
\end{equation*}
where $E,E^*$ are the complex solutions given by~\citet{Abramowitz}. As noted by \citet{Narayan1987}, the functions $E$ and $E^*$ represent radially outward and inward propagating waves, respectively. The boundary conditions on $u$ are transformed into the conditions 
\begin{equation*}
    2a\int_{-1}^{\zeta(1)} \sqrt{\xi^2-1}~d\xi = \int_{\mu_1}^{1} \sqrt{Q} ~d\xi,
\end{equation*}
and
\begin{equation*}
    2a\int_{1}^{\zeta(r_2/r_1)} \sqrt{\xi^2-1}~d\xi = \int_{\mu_2}^{r_2/r_1} \sqrt{Q} ~d\xi 
\end{equation*}
To satisfy both boundary conditions, we require
$v(\zeta(1))=0=v(\zeta(r_2/r_1))=0$. Or,
\begin{equation}
    \frac{E\left(a,2\sqrt{a}\zeta(1)\right)}{E^*\left(a,2\sqrt{a}\zeta(1)\right)} = -\frac{E^*\left(a,2\sqrt{a}\zeta(r_2/r_1)\right)}{E\left(a,2\sqrt{a}\zeta(r_2/r_1)\right)}
\end{equation}
Since $\zeta(1)$ has $\arg(\zeta)\approx \pi $, we can use the connection formulas $E(a,-z)=-ie^{a\pi}E(a,z)+i\sqrt{1+e^{2\pi a}}E^*(a,z)$ and $E^*(a,-z)=ie^{a\pi}E^*(a,z)-i\sqrt{1+e^{2\pi a}}E(a,z)$. Thus
\begin{equation}
    \frac{e^{a\pi}E\left(a,-2\sqrt{a}\zeta(1)\right)-\sqrt{1+e^{2\pi a}}E^*\left(a,-2\sqrt{a}\zeta(1)\right)}{e^{a\pi}E^*\left(a,-2\sqrt{a}\zeta(1)\right)-\sqrt{1+e^{2\pi a}}E\left(a,-2\sqrt{a}\zeta(1)\right)} = \frac{E^*\left(a,2\sqrt{a}\zeta(r_2/r_1)\right)}{E\left(a,2\sqrt{a}\zeta(r_2/r_1)\right)} .
    \label{eq: par_cyl_disp}
\end{equation}
This facilitates the use of the standard asymptotic relations given for $E(a,z)$ for large $|z|$ when $|\arg{z}| < \pi/4  $. Applying this formula requires careful additional levels of approximation in general. It is worth noting the physical content of this expression. The right-hand side represents the approximate ratio of the radially inward propagating wave to the radially outward propagating wave in the region to the right of the second turning point, $\mu_2$, evaluated at the location of the outer wall. For $s>\mu_2, \zeta> 1$, the radially outward propagation is also the direction \emph{away} from the turning point. On the left-hand side, the roles of $E$ and $E^*$ are reversed. $E$ now propagates towards the turning point, and $E^*$ propagates away. The factors $\sqrt{1+e^{2\pi a}}$ and $e^{a\pi}$ relate the magnitudes of the reflected and transmitted waves on either side of the forbidden region $\zeta^2 <1$. To see this more clearly, consider the effect of removing the outer boundary to $r_2/r_1 \rightarrow \infty $, and demanding that the solution for $s>\mu_2, \zeta > 1$ represent only radially outward propagating waves. Then we would require
\begin{equation*}
    \frac{E\left(a,-2\sqrt{a}\zeta(1)\right)}{E^*\left(a,-2\sqrt{a}\zeta(1)\right)}=\frac{\sqrt{1+e^{2\pi a}}}{e^{a\pi}}.
\end{equation*}
Evidently, the ratio of the magnitude the reflected wave to the transmitted wave is $\sqrt{1+e^{2\pi a}}/e^{a\pi}$. The radially outward wave suffers an over-reflection near the corotation point. In the following we show that it is precisely this amplitude mismatch that gives rise to a growing mode in this case. We can get an approximate dispersion relation by using the large argument asymptotic expressions for $E$ and $E^*$. 
\begin{equation}
    \frac{\exp(ia\zeta^2-ia\log(-2\sqrt{a}\zeta)+i\pi/4)}{\exp(-ia\zeta^2+ia\log(-2\sqrt{a}\zeta)-i\pi/4)} \approx \frac{\sqrt{1+e^{2\pi a}}}{e^{a\pi}},
\end{equation}
yielding
\begin{equation}
      a\zeta^2(1) - a\log(-2\sqrt{a}\zeta(1)) + \frac{\pi}{4} \approx -\frac{i}{2}\log(\frac{\sqrt{1+e^{2\pi a}}}{e^{a\pi}}) + n\pi,
\end{equation}
for some integer $n$. When $ |\zeta|\gg 1, \zeta<0$, we have $a\zeta^2 \sim a - \int_{\mu_1}^s \sqrt{Q}~d\xi + a\log(-\zeta)$, so
\begin{equation}
     \int_{1}^{\mu_1}\sqrt{Q}~ds \approx  a\log(2\sqrt{a}) -a -\frac{i}{2}\log(\frac{\sqrt{1+e^{2\pi a}}}{e^{a\pi}}) +\left(n-\frac{1}{4}\right)\pi.
\end{equation}
Note that $a,\mu_1,\mu_2$ and $Q$ all depend on $\omega$. In the region away from the turning point $\int_1^{\mu_1} \sqrt{Q} \sim (r_1/d_i)\int_{1}^{\mu_1} s^2|\obar|/mV_A$. So we have,
\begin{equation}
    \int_1^{\mu_1}   m\sqrt{s}-s^2\omega ~ds \approx \frac{md_iV_A}{r_1} \left(a\log(2\sqrt{a}) -a -\frac{i}{2}\log(\frac{\sqrt{1+e^{2\pi a}}}{e^{a\pi}}) +\left(n-\frac{1}{4}\right)\pi\right).
\end{equation}
From this expression we can extract the approximate growth rate, $\gamma$ to be
\begin{equation}
    \gamma \sim \frac{3md_iV_A}{2r_1}\frac{\log(\sqrt{1+e^{-2\pi a}})}{\mu_1^3-1}. \label{eq: gamma_one_wall}
\end{equation}
Recalling the definition of $\mu_1$, we have $\mu_1^3 = (m^2/4\omega^2)\left(1+\sqrt{1-4\omega d_iV_A\sqrt{m^2-1/4}}\right)^2$.
\par Returning to the two-wall case, we can apply the asymptotic relations for the parabolic cylinder functions to equation \eqref{eq: par_cyl_disp}
\begin{equation}
     \exp(2i\vartheta_1) - \exp(-2i\vartheta_2) = \sqrt{1+e^{-2\pi a}}\left(1 - \exp(2i(\vartheta_1-\vartheta_2))\right).
\end{equation}
 where $\vartheta_1 \equiv a\zeta^2(1)-a\log(-2\sqrt{a}\zeta(1))+\pi/4$ and $\vartheta_2\equiv a\zeta^2(r_2/r_1) - a\log(2\sqrt{a}\zeta(r_2/r_1)) + \pi/4$.
 
 \par Whether the domain is bounded by one wall or two, the growth rate depends on the area contained by the forbidden region, $Q<0$, via the parameter $a$. For a given  $\omega$, we can approximate $Q$ on the interval $\mu_1\leq s \leq \mu_2$ as $Q \approx 4|Q_{m}|(s-\mu_1)(s-\mu_2)/(\mu_2-\mu_1)^2$, where $Q_m$ is the minimum value of $Q$ on the interval. $|Q_m|$ does not depend on $V_A$, and decreases slowly as $\omega$ decreases. The `potential well' becomes shallower as it moves closer to the outer wall. This is essentially an effect of the Keplerian rotation profile and cylindrical geometry. The width of the well, however, depends almost entirely on the quantity $d_i V_A$. Using this estimate for $Q$ in the region between the two turning points, we have $a\sim \sqrt{|Q_m|}(\mu_2-\mu_1)/4$. As $V_A\rightarrow 0$ the width of the interval vanishes and $a\rightarrow 0$.

\bibliography{refs}
\bibliographystyle{aasjournalv7}

\end{document}